\documentclass{article}

\usepackage{arxiv}

\usepackage[utf8]{inputenc} 
\usepackage[T1]{fontenc}    
\usepackage[hidelinks]{hyperref}       
\usepackage{url}            
\usepackage{booktabs}       
\usepackage{amsfonts}       
\usepackage{nicefrac}       
\usepackage{microtype}      
\usepackage{graphicx}
\usepackage{natbib}
\usepackage{doi}
\hypersetup{
  colorlinks   = true, 
  urlcolor     = black, 
  linkcolor    = black, 
  citecolor   = blue 
}

\title{Emerging AI-based weather prediction models as downscaling tools}

\author{ \href{https://orcid.org/0000-0002-3365-8146}{\includegraphics[scale=0.06]{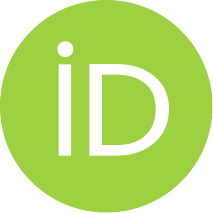}\hspace{1mm}Nikolay Koldunov}\thanks{Use footnote for providing further
		information about author (webpage, alternative
		address)---\emph{not} for acknowledging funding agencies.} \\
	Alfred Wegener Institute\\
	  Helmholtz Centre for Polar\\and Marine Research (AWI)\\
	\texttt{nikolay.koldunov@awi.de} \\
	\And
	\href{https://orcid.org/0000-0002-5468-575X}{\includegraphics[scale=0.06]{orcid.pdf}\hspace{1mm}Thomas Rackow} \\
	European Centre for Medium-Range\\ Weather Forecasts (ECMWF)\\
	\texttt{thomas.rackow@ecmwf.int} \\
\And
        \href{https://orcid.org/0000-0002-2740-6815}{\includegraphics[scale=0.06]{orcid.pdf}\hspace{1mm} Christian Lessig} \\
	European Centre for Medium-Range\\ Weather Forecasts (ECMWF)\\
	\texttt{christian.lessig@ecmwf.int} \	
 \And
	\href{https://orcid.org/0000-0001-8098-182X}{\includegraphics[scale=0.06]{orcid.pdf}\hspace{1mm}Sergey Danilov} \\
	Alfred Wegener Institute\\
	  Helmholtz Centre for Polar\\and Marine Research (AWI)\\
	\texttt{sergey.danilov@ecmwf.int} \\
        \And
	\href{https://orcid.org/0000-0000-0000-0000}{\includegraphics[scale=0.06]{orcid.pdf}\hspace{1mm}Suvarchal K. Cheedela} \\
	Alfred Wegener Institute\\
	  Helmholtz Centre for Polar\\and Marine Research (AWI)\\
	\texttt{suvarchal.cheedela@awi.de} \\
        \And
	\href{https://orcid.org/0000-0001-8579-6068}{\includegraphics[scale=0.06]{orcid.pdf}\hspace{1mm}Dmitry Sidorenko} \\
	Alfred Wegener Institute\\
	  Helmholtz Centre for Polar\\and Marine Research (AWI)\\
	\texttt{dmitry.sidorenko@ecmwf.int} \\
        \And
	\href{https://orcid.org/0000-0002-1215-3288}{\includegraphics[scale=0.06]{orcid.pdf}\hspace{1mm}Irina Sandu} \\
	European Centre for Medium-Range\\ Weather Forecasts (ECMWF)\\
	\texttt{irina.sandu@ecmwf.int} \\
         	\And
	\href{https://orcid.org/0000-0002-2651-1293}{\includegraphics[scale=0.06]{orcid.pdf}\hspace{1mm} Thomas Jung} \\
	Alfred Wegener Institute\\
	  Helmholtz Centre for Polar\\and Marine Research (AWI)\\
	\texttt{thomas.jung@awi.de} \\
}

\renewcommand{\shorttitle}{AI-NWP as downscaling tools}

\hypersetup{
pdftitle={ainwp_downscaling},
pdfsubject={physics.ao-ph},
pdfauthor={Nikolay Koldunov et al.},
pdfkeywords={climate, climate modelling, downscaling, machine learning},
}

\begin{document}
\maketitle

\begin{abstract}
The demand for high-resolution information on climate change is critical for accurate projections and decision-making. Presently, this need is addressed through high-resolution climate models or downscaling. High-resolution models are computationally demanding and creating ensemble simulations with them is typically prohibitively expensive. Downscaling methods are more affordable but are typically limited to small regions. This study proposes the use of existing AI-based numerical weather prediction systems (AI-NWP) to perform global downscaling of climate information from low-resolution climate models.

Our results demonstrate that AI-NWP initalized from low-resolution initial conditions can develop detailed forecasts closely resembling the resolution of the training data using a one day lead time. We constructed year-long atmospheric fields using AI-NWP forecasts initialized from smoothed ERA5 and low-resolution CMIP6 models. Our analysis for 2-metre temperature indicates that AI-NWP can generate high-quality, long-term datasets and potentially perform bias correction, bringing climate model outputs closer to observed data.

The study highlights the potential for off-the-shelf AI-NWP to enhance climate data downscaling, offering a simple and computationally efficient alternative to traditional downscaling techniques. The downscaled data can be used either directly for localized climate information or as boundary conditions for further dynamical downscaling.
\end{abstract}

\keywords{climate \and climate modelling \and downscaling \and machine learning}

\section{Introduction}

In the last two years, several data-driven AI-based numerical weather prediction systems (AI-NWP) for medium-range global weather forecasting have been developed. Notable examples include Pangu-Weather \citep{bi2023accurate}, GraphCast \citep{lam2023learning}, FourCastNet \citep{pathak2022fourcastnet}, FuXi \citep{chen2023fuxi} and AIFS \citep{lang2024aifs}. 
These models today outperform the best conventional, equation-based models in deterministic skill for most variables and metrics \citep{BenBouallegue2023}.
The models are trained on the ERA5 reanalysis \citep{hersbach2020era5} that combines a large number of observations into a best state estimate of the atmosphere. It has resolution of about 31\,km. 
When used to generate medium-range  forecasts, the AI-NWP models use either ERA5 data, operational ECMWF forecasts, or analysis data as initial conditions. 

In this work, we exploit that AI-NWP models are trained to provide predictions that match ERA5 both in terms of its resolution and its data distribution and that they have demonstrated a surprising robustness to novel initial conditions \citep{Hakim2023}.
Based on this, we hypothesize that AI-NWP models initialised with low-resolution initial conditions will produce output with a higher level of detail matching their training data. Through this, they will effectively perform automatic downscaling without being trained for it, an approach known as zero-shot. Since the ERA5 reanalysis is observationally well-constrained, they potentially can also bias correct model output, e.g. from low resolution biased climate simulations.

Downscaling of data from low-resolution climate models is a well-developed field \citep{giorgi2019rcm} and currently forms the basis for regional climate assessments, such as those in the IPCC reports \citep{ranasinghe2021ipcc}. This is usually achieved with regional climate models (RCMs), which use low-resolution global climate models as boundary conditions and run simulations for smaller regions, often in atmosphere-only mode, although coupled regional climate simulations are also well developed (e.g. \cite{ruti2016med}). There are also statistical downscaling techniques, which employ statistical relationships between low-resolution inputs and high-resolution models or observations to downscale model results (\cite{fowler2007stat, ekstrom2015stat}). Downscaled climate data are fundamental for climate service products, as users require information on a local scale (e.g. \cite{van2018match, koldunov2016identifying, damm2020market}). The most prominent coordinated initiative in regional downscaling is the Coordinated Regional Climate Downscaling {EX}periment (CORDEX, \cite{giorgi2015cordex}).

\newcommand{\citepeg}[1]{\citep[e.g.][]{#1}}

Neural networks have been used extensively for climate and weather data downscaling in recent years \citepeg{bano2020configuration, rampal2022high, hohlein2020comparative}. Most of the applications target weather prediction use cases, regional domains and use low/high resolution pairs to train the neural networks \citep{stengel2020adversarial, harris2022generative}. An interesting approach is to emulate dynamical downscaling done with RCMs, using low resolution climate simulations \citep{tomasi2024ai, mardani2023residual} or reanalysis \citep{wang2021fast} as low-resolution data, and RCM simulations as the target high-resolution data. Foundation models for weather and climate can also be employed for downscaling tasks, and there are examples for regional \citep{lessig2023atmorep} and global climate \citep{nguyen2023climax}.

In this work we will first evaluate the approach of downscaling with AI-NWP by applying it to low-resolution data given by smoothed ERA5 fields. We then study whether the AI-NWP models produce physically plausible output fields that have more fine-scale detail than the input. Next, we will assess whether it is feasible to construct a high-quality, long-term (year-long) record of atmospheric fields by combining the results from a sequence of forecasts initiated from low-resolution data. Finally, we will use model output of low-resolution CMIP6 models as initial conditions for a AI-NWP moddel for several years of historical and future scenario simulations.

The aim of this paper is to demonstrate that this approach is viable in principle. Therefore, we will focus primarily on surface temperature, one of the main climate parameters with significant impact, and use a limited number of years as examples. More detailed experiments are planned for the near future.

\section{Methods}
\subsection{Pangu-Weather}

We use Pangu-Weather \citep{bi2023accurate} as AI-NWP model in this study. Pangu-Weather uses a 3D Earth-specific transformer architecture, designed to capture complex atmospheric interactions across multiple pressure levels and surface variables. The model is trained on 39 years of ERA5 reanalysis data (1979 to 2017), encompassing global atmospheric and surface variables at high temporal and spatial resolution, corresponding approximately to 31\,km. The input variables for the Pangu-Weather model include geopotential, specific humidity, temperature, and the u- and v-components of wind on 13 pressure levels (50 hPa, 100 hPa, 150 hPa, 200 hPa, 250 hPa, 300 hPa, 400 hPa, 500 hPa, 600 hPa, 700 hPa, 850 hPa, 925 hPa, and 1,000 hPa), as well as surface variables such as 2-metre temperature, the u- and v-component of 10-meter wind, and mean sea level pressure. Pangu-Weather outperforms the operational Integrated Forecasting System (IFS) of the European Centre for Medium-Range Weather Forecasts (ECMWF) in deterministic forecasts for many relevant metrics. It also demonstrates superior performance with respect to tropical cyclone tracks providing more accurate and computationally efficient predictions.

Pangu-Weather is used as distributed through the \texttt{ai-models} package \citep{aimodels}, created by ECMWF to run different AI-based weather forecasting models from a similar interface. Runs were performed at the German Climate Supercomputing Centre (DKRZ), on the Levante HPC's Graphical Processing Unit (GPU) nodes equipped with 4 NVIDIA A100-SXM4 GPUs. One 10-day forecast takes about 2 minutes of wall clock time on a single GPU and produces a 5.7GB output file (6-hourly output frequency).

\subsection{ERA5 and climate data preparation}

As mentioned before, Pangu-Weather is called through the \texttt{ai-models} package, which requires data in GRIB format, interpolated on a regular 0.25$^\circ$ degree regular grid. ERA5 data were interpolated from the original reduced Gaussian grid to a regular grid using the \texttt{cdo} package \citep{schulzweida_2023_10020800} with bi-linear interpolation (based on version implemented through YAC \citep{hanke2016yac}). As the GRIB files produced by \texttt{cdo} are not compatible with the input expected by \texttt{ai-models}, the interpolated data were converted to GRIB format using the \texttt{climetlab} package \citep{climetlab}. To smooth the initial conditions, we applied Gaussian smoothing using the \texttt{scipy.ndimage.gaussian\_filter} function from the \texttt{scipy} package \citep{2020SciPy-NMeth}, with various \texttt{sigma} values.

As low-resolution climate simulations we use CMIP6 data, namely AWI-ESM-1-1-LR \citep{awiesmdata} based on the AWI-CM model \citep{sidorenko2015awicm, rackow2018awicm, semmler2020simulations}, and MPI-ESM1-2-LR \citep{mpi_hist, mpi_scenario} based on MPI-ESM \citep{giorgetta2013climate}. Both models share the same atmospheric component ECHAM6.3 \citep{stevens2013atmospheric}, configured with a T63 Gaussian grid with 192\,longitudes, 96\,latitudes, and 47\,vertical levels. The models however use different ocean components: FESOM1.4 \citep{wang2014finite} and MPI-OM \citep{jungclaus2013characteristics}. This leads to different climate sensitivities in their higher-resolution versions \citep{semmler2021ocean}. The atmospheric resolution of those models is approximately 250 km. This resolution is typical for CMIP3, whereas CMIP6 generally features resolutions of around 100\,km \citep{haarsma2016high}. We will explore low-resolution models with the expectation that results for higher-resolution models will be similar or better, since those are closer to the target resolution of AI models. 

Scenario simulations for AWI-ESM-1-1-LR are not available, so we only use projections from MPI-ESM1-2-LR SSP585 scenario \citep{mpi_scenario}.  Data were accessed from the \href{https://esgf-data.dkrz.de/projects/esgf-dkrz/}{ESGF node at DKRZ }. We interpolated these data to a 0.25$^\circ$ regular grid from the original CMORised netCDF files using \texttt{cdo}'s bi-linear interpolation. The $u$ and $v$ fields on pressure levels required an additional step because the necessary pressure levels were not available in the original CMIP6 output. Therefore, we interpolated the data from model levels, which was also done using \texttt{cdo}.

\subsection{Processing Software}
In addition to already mentioned software, data processing, analysis, and visualisation were performed with the following python packages: xarray \citep{hoyer2017xarray}, dask \citep{dask}, zarr \citep{zarr}, matplotlib \citep{matplotlib}, cartopy \citep{cartopy}, regionmask \citep{regionmask}.

\section{Results}

\subsection{Initialisation with smoothed ERA5 fields}

In this section, we assess whether Pangu-Weather produces meaningful outputs when initialized from smoothed, i.e., coarsened, data, whether we can combine the resulting forecasts into a long-term time series of meteorological fields, and how the statistics of these fields compare to the original ERA5 data.

We initialize the Pangu-Weather using the original ERA5 data, interpolated to a regular grid (AI orig experiment), and data smoothed with different $\sigma$ values (standard deviations for the Gaussian kernel): $\sigma=2$ (AI 2$\sigma$), $\sigma=5$ (AI 5$\sigma$), and $\sigma=10$ (AI 10$\sigma$). The approximation of the equivalent resolution after smoothing can be done by using full width at half maximum of the Gaussian function by multiplying the initial resolution by $2.355 \times \sigma$. Given that the 0.25$^\circ$ regular grid corresponds to approximately 25 km resolution at the equator, AI 2$\sigma$ approximates a 100 km resolution, AI 5$\sigma$ corresponds to about 300 km resolution, and AI 10$\sigma$ to 500 km resolution. These estimations are approximate and do not account for the original ERA5 resolution of approximately 31\,km, the variation in horizontal resolution of regular grids with latitude, or the inability of a 500 km resolution model to resolve certain features, such as atmospheric front positions, that are present in the smoothed data.

\begin{figure}
  \centering
  \includegraphics[width=0.9\textwidth]{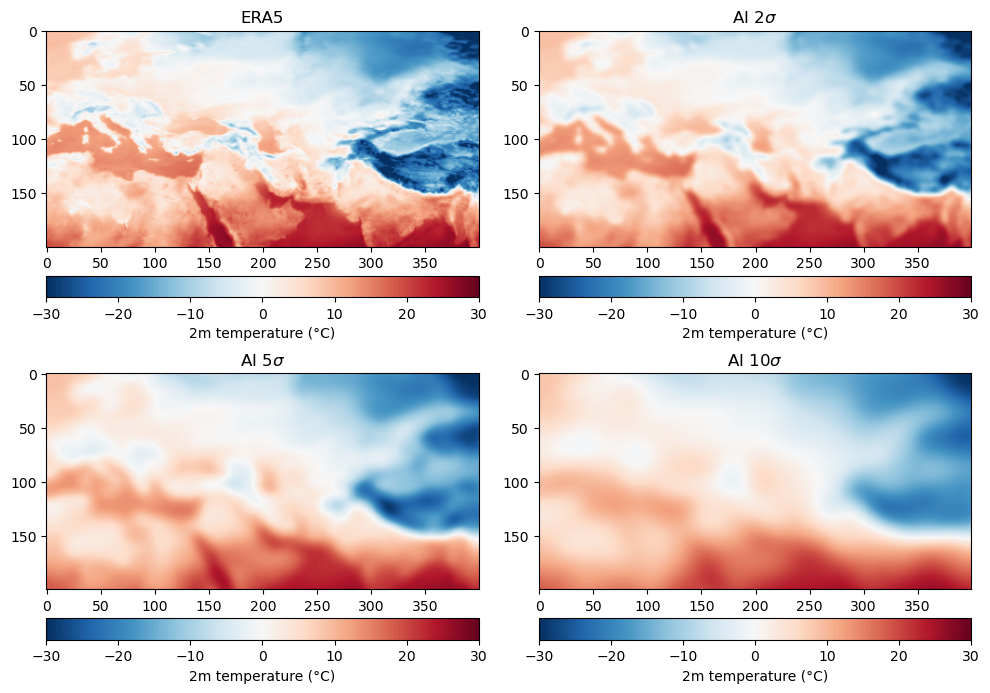}
  \caption{The 2-metre temperature used as part of the initial conditions (2020-01-01T00:00:00) with several levels of Gaussian smoothing. All other initial fields were smoothed analogously to emulate the fields generated by atmospheric models with different horizontal resolutions.}
  \label{fig:fig1}
\end{figure}

The initial 2-metre temperature fields provided to AI-NWP are shown in Figure \ref{fig:fig1}. This figure displays parts of Europe, Asia, and Africa for the original ERA5 data (used to initialize the AI orig experiment) and data with varying levels of smoothing. For AI 2$\sigma$ and AI 5$\sigma$, a considerable amount of detail remains visible. However, AI 10$\sigma$ retains primarily large-scale features. Figure \ref{fig:A1} presents a global view of the same fields.

\begin{figure}
  \centering
  \includegraphics[width=0.9\textwidth]{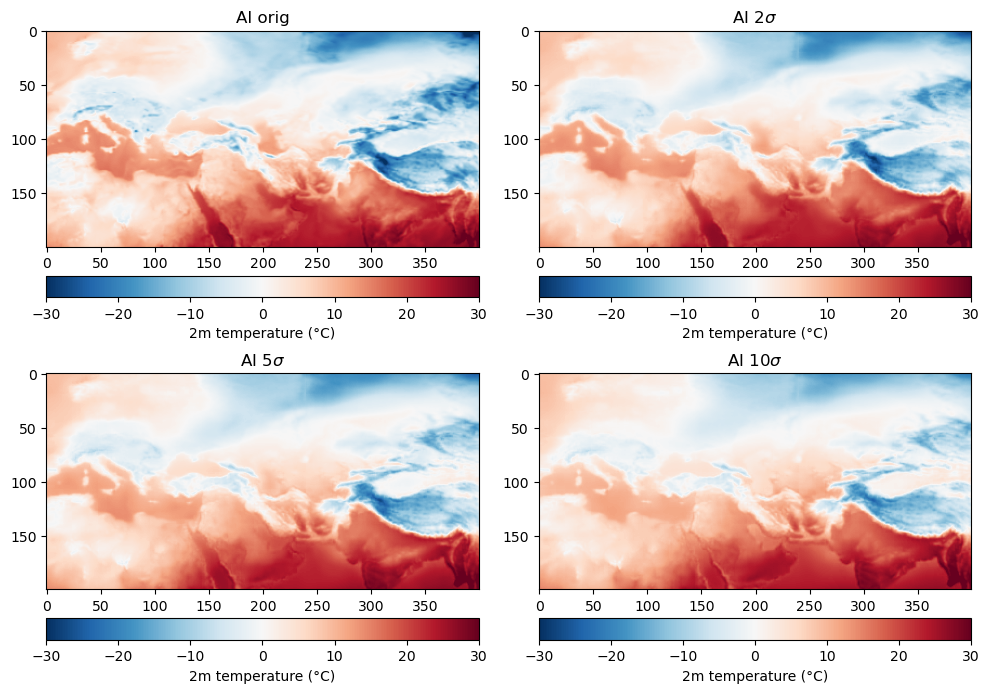}
  \caption{The 2-metre temperature for a day forecast (2020-01-02T00:00:00) using the AI-NWP for simulations with varying smoothing of initial conditions. After one day, the AI-NWP enhances the level of detail to a rate comparable to the fields it was originally trained on (ERA5).}
  \label{fig:fig2}
\end{figure}

After one day of forecast, the 2-metre temperature fields initialized with different levels of smoothing converge significantly in terms of detail (Figure \ref{fig:fig2}). The contrast between initial fields of AI 5$\sigma$ and 10$\sigma$ and the AI-NWP forecast is noticeable. Numerous small-scale details emerge, distinctions between land and ocean areas become evident, and mountain ridges and large valleys become visible in the temperature fields. Overall, the level of detail resembles that of the original ERA5 fields. Relatively high level of detail appears as early as the first forecast step (Figure \ref{fig:A2}).

Next we examine how initialization from smoothed initial conditions affects the statistics of the resulting temperature fields. We conducted a 10-day forecast for each day of the year 2020 using AI-NWP, initialized (at 00:00) from both the original (interpolated to a regular grid) and smoothed ERA5 data. We then constructed two annual datasets of atmospheric 2-metre temperature fields: one consisting of all first days (1-day), and the second dataset concatenating all second days (2-day) of the individual forecasts. For the 1-day dataset, the first time step is the initial condition, and the lead hours 6, 12, and 18 are from AI-NWP forecasts. For the 2-day dataset, all time steps are from AI-NWP forecasts (lead times of 24, 30, 36, and 42 hours).

\begin{figure}[htbp]
  \centering
  \makebox[\textwidth][c]{%
    \includegraphics[width=1.1\textwidth]{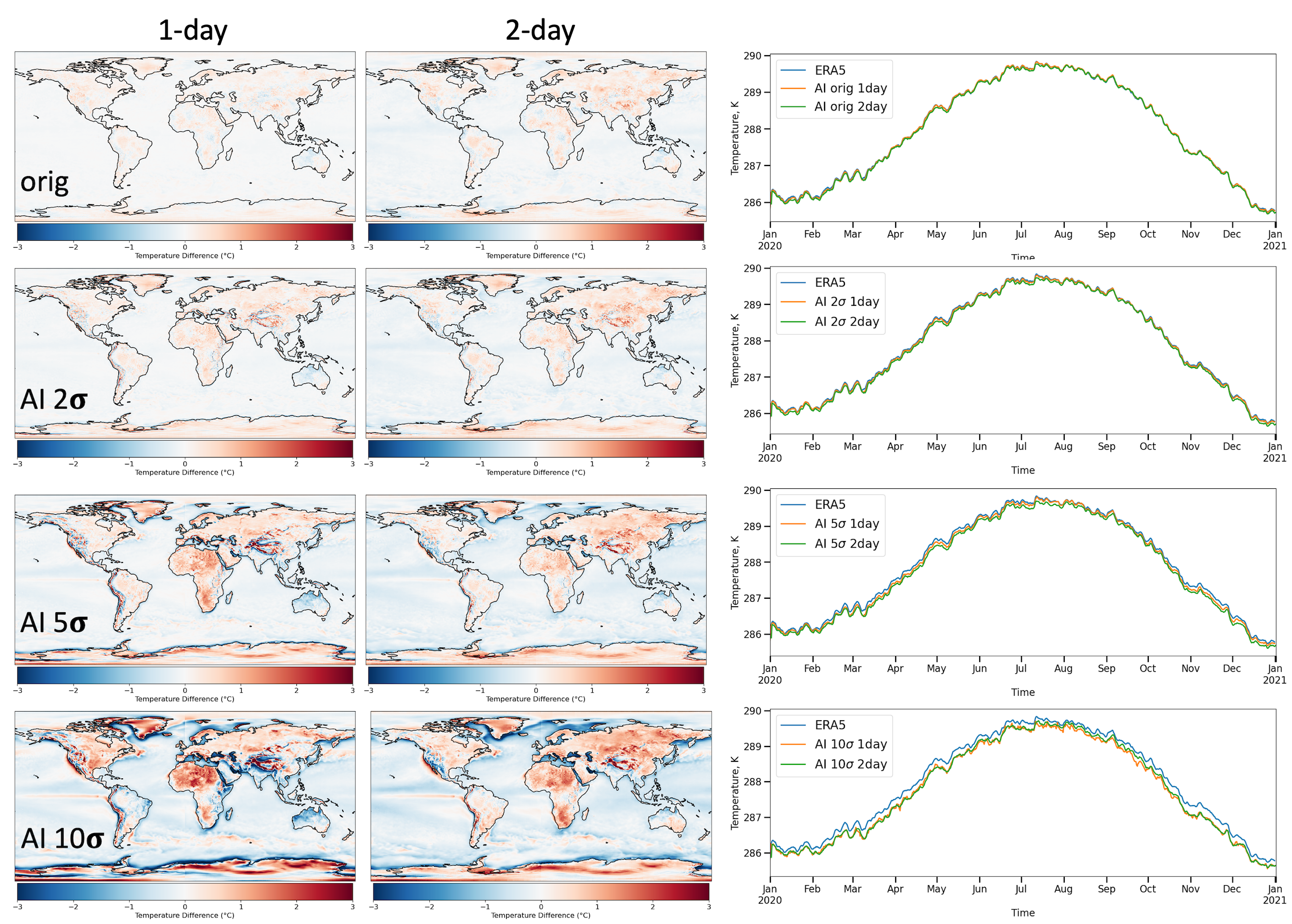}
  }
  \caption{Annual mean 2-metre temperature difference relative to ERA5 (maps) and time series of global mean 2-metre temperature (graphs, 6-hour values, daily rolling mean). Values are compared for time series constructed from the first (1-day) and second (2-day) days of the AI-NWP forecast with varying levels of initial condition smoothing (original and multiple values of $\sigma$, the standard deviation for the Gaussian kernel). The level of $\sigma = 5$ corresponds to fields from 300 km resolution models and can serve as a reference for low-resolution climate model fields. Constructing 6-hour time series from the first and second days of the forecast is effective, reproducing ERA5 properties well when initialized from original ERA5 data. Smoothing initial conditions tends to push the model towards a colder state. The main source of the cold bias appears to be the temperature over the ocean.}
  \label{fig:fig3}
\end{figure}

\begin{table}[h]
\centering
\begin{tabular}{|l|c|c|}
\hline
Smoothing & 1 day, $^{\circ}$C & 2 day, $^{\circ}$C \\
\hline
AI orig. & -0.004 & -0.032 \\
AI 2$\sigma$ & -0.023 & -0.069 \\
AI 5$\sigma$ & -0.065 & -0.129 \\
AI 10$\sigma$ & -0.196 & -0.177 \\
\hline
\end{tabular}
\caption{Mean differences in global mean 2-metre temperature time series relative to ERA5 (refer to Fig \ref{fig:A3}) for the time series of the differences).}
\label{table:smoothing}
\end{table}

The global mean 2-metre temperature time series computed from the 1-day and 2-day datasets closely match the ERA5 data for AI-NWP forecasts initialized from unsmoothed initial conditions (Figure \ref{fig:fig3}, first row, right). The 1-day line is nearly indistinguishable from the original ERA5 line, with a mean difference of -0.004$^\circ$C (Table \ref{table:smoothing}). This indicates that the approach of creating datasets of meteorological parameters by "filling" the gaps between ERA5 daily time steps with AI-NWP-generated data performs well. The 2-day line, although still close to the ERA5, exhibits larger differences, indicating some degradation of the results, with the mean difference becoming -0.032$^\circ$C (Table \ref{table:smoothing}). The spatial distribution of temperature differences for the AI orig. experiment (Figure \ref{fig:A3}, first row, left and middle) shows relatively small biases for both the 1-day and 2-day datasets, with predominantly negative bias over the ocean and predominantly positive bias over the land.

The differences between ERA5 data and AI-NWP experiments with smoothed initial conditions increase with the level of smoothing (Figure \ref{fig:fig3}). Notably, for AI 5$\sigma$ and AI 10$\sigma$, some spatial biases over land for 2-day data are smaller than those for 1-day data, particularly over regions such as Africa and Australia. This suggests an adjustment from the smoothed initial conditions, which is evident in the mean 2-metre temperature RMSE values across different experiments (Figure \ref{fig:A4}). This adjustment is also reflected in the larger mean differences in global 2-metre temperature for the 2-day compared to the 1-day forecast in the AI 10$\sigma$ experiment (Table \ref{table:smoothing}). The overall cool bias is primarily due to a uniform cooling effect over the ocean.

The results presented in this section indicate that AI-NWP can be initialized from smoothed initial conditions, and the resulting forecasts will have increased levels of detail, matching the resolution of the data on which AI-NWP was trained after a few forecasting steps. Year-long 2-metre temperature datasets, constructed from the first and second days of AI-NWP forecasts, represent well the temporal evolution and spatial distribution of the original data. The quality of the results decreases with increased smoothing of initial conditions, but for AI 5$\sigma$, which is approximately equivalent to a 300 km resolution, the quality remains acceptable (for example well below typical biases of CMIP6 models \citep{fan2020global}). The data for the second day of the forecast (2-day) exhibit smaller biases over land in the AI 5$\sigma$ experiment. Therefore, for further experiments with climate models, we will use the 2-day data.

\subsection{Initialisation with historical climate model data}

\begin{figure}
  \centering
  \includegraphics[width=0.95\textwidth]{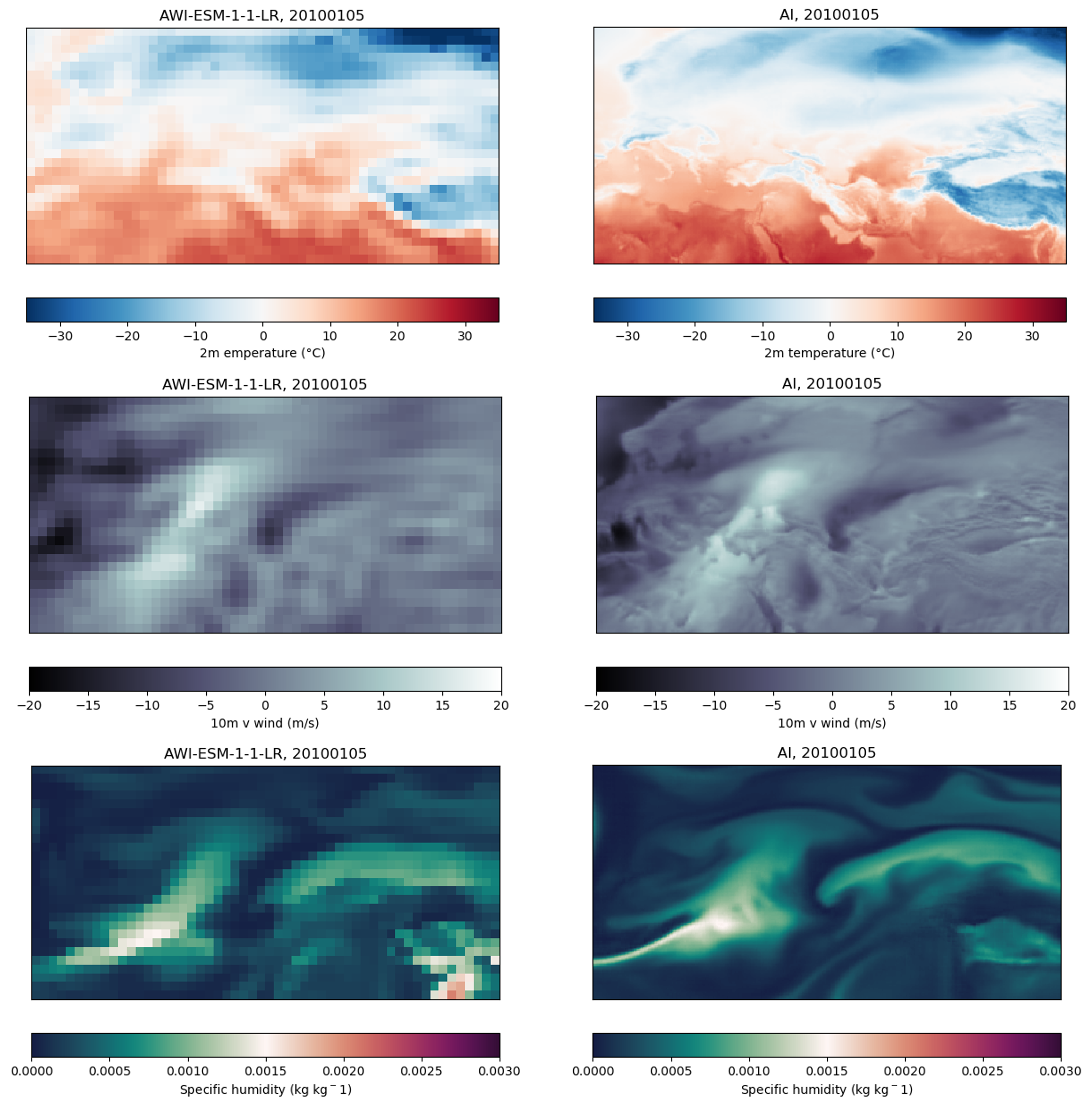}
  \caption{Snapshot from 2010-01-05 comparing original fields from a low-resolution climate model (left) and the result of an AI-NWP forecast initialized 2 days prior from climate model data (right). The region displayed includes parts of Europe, Africa, and Asia.}
  \label{fig:fig6}
\end{figure}

Next, we use low-resolution climate models (AWI-CM-LR and MPI-ESM1-2-LR) to initialize AI-NWP forecasts at 00:00 UTC each day for one model year (2010) from historical experiments. The AI-NWP forecasts rapidly developed a level of detail comparable to the ERA5 data, as demonstrated for 2-metre temperature, 10-meter v wind, and specific humidity fields over regions in Europe, Africa, and Asia for AI-ESM-1-1-LR (Fig. \ref{fig:fig6}). Similarly to the ERA5 case, there is an improved contrast between ocean and land, as well as more detailed representation of orographic features. Global fields also indicate that the large-scale patterns from the input are largely reproduced in AI-NWP, but with a greater level of detail, which appears plausible from a dynamical perspective (Figure \ref{fig:A5}).

\begin{figure}
  \centering
  \includegraphics[width=0.95\textwidth]{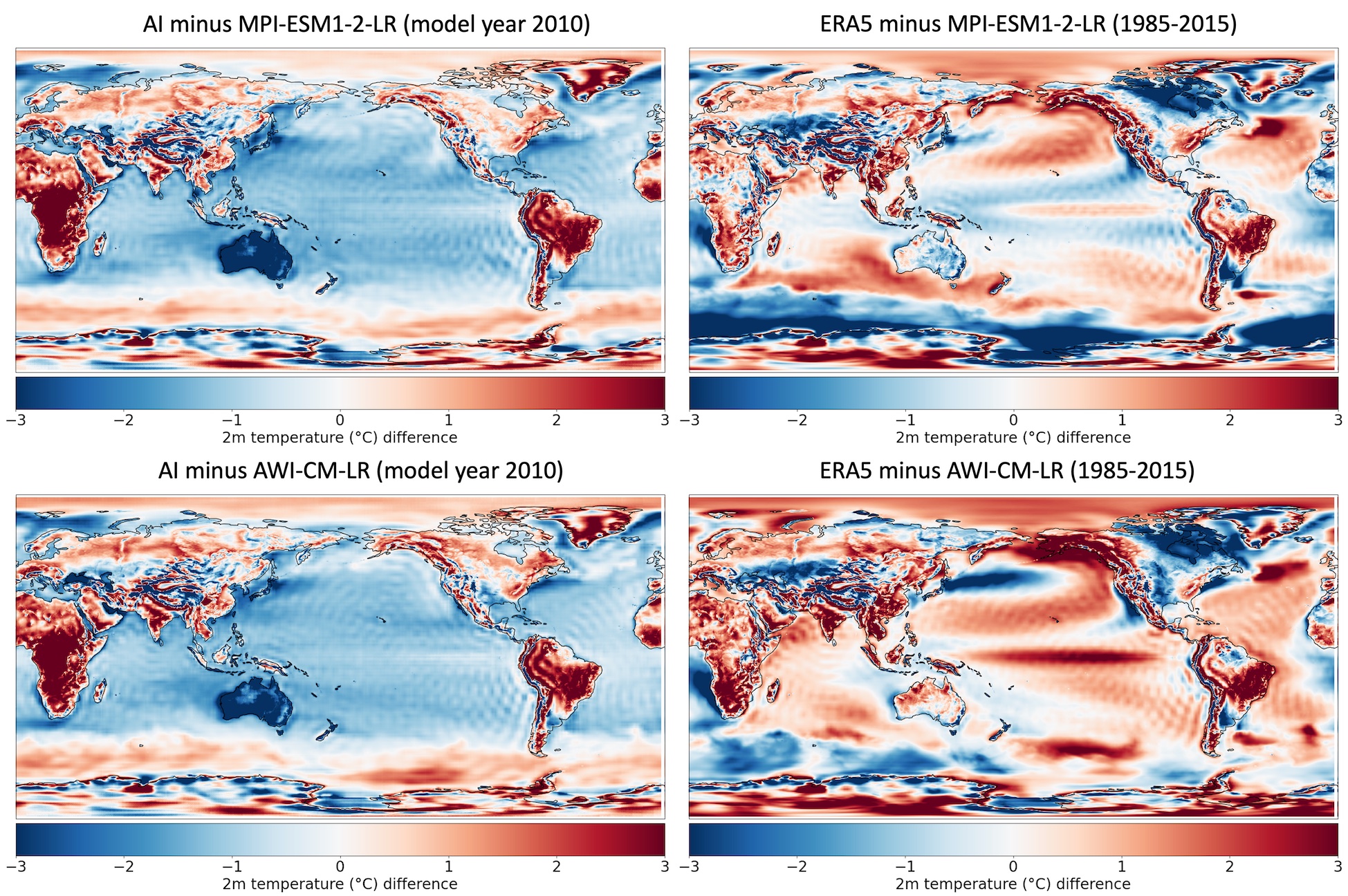}
  \caption{Difference in annual mean 2-metre temperature. (left) Between AI-NWP generated data and two CMIP6 climate models for model year 2010 (historical simulations). (right) Between ERA5 and climate models for the years 1985-2015 (historical simulations). The AI-NWP time series is derived from lead hours 24, 30, 36, and 42 (2-day dataset) of AI-NWP forecasts initiated from CMIP6 model fields. Two low-resolution (250km) models are used: AWI-CM-LR and MPI-ESM1-2-LR. Both models share the same ECHAM atmosphere but use different ocean components. This figure illustrates the bias of climate models relative to ERA5 and assesses whether the AI-NWP adjusts the climate model towards more realistic values. The AI-NWP model improves realism over large land areas, but problematic regions include the oceans, Australia, and northeastern North America.}
  \label{fig:fig7}
\end{figure}

Utilizing the same methodology as in the previous section for ERA5, we constructed one-year-long datasets from the second days of AI-NWP forecasts performed daily over the course of a year. The annual mean temperature difference between these datasets (AI) and the original climate model data from the corresponding time period (interpolated to a 0.25° grid) indicates the average direction of change that AI-NWP does for the temperature fields (Figure \ref{fig:fig7}, left column). The differences for both climate models are similar in terms of magnitude and spatial distribution. Over most ocean areas, the temperature is lower in AI-NWP, except in the Southern Ocean, where it is warmer. Over land areas, AI-NWP data generally show higher temperatures, except for a significant negative difference over Australia.

To contextualize these differences, we compare the mean 2-metre temperature from 30 years of ERA5 data with climate model simulations (Figure \ref{fig:fig7}, right column). Note, that we subtract model data from ERA5 rather than the reverse, to ensure comparability with the plot in the left column. In this way, similar colors indicate that the AI-NWP based dataset becomes closer to ERA5, and vice versa. The spatial distribution of differences relative to ERA5 for both climate models over land is similar, suggesting comparable model biases over land in historical simulations. When compared to the differences between AI-NWP data and climate models for one year, the spatial distribution, sign, and magnitude of the differences over land are quite similar. This suggests that the AI-NWP forecast moves the solution closer to ERA5, effectively performing a bias correction. Notable exceptions are Australia and northeastern North America, where the AI-NWP forecast is too cold. Additionally, temperatures over the ocean in AI-NWP-based data are too cold. Note that despite obvious and rather strong differences over the ocean in right plots for Figure \ref{fig:fig7}, the difference in the left plot is remarkably similar, which is one more indication that solutions are moved in similar way. It should be also noted that for the AI-NWP based data, only one year is analyzed here as a demonstrator of the AI-NWP downscaling capability. A fairer comparison with the 30-year dataset will involve more AI-generated annual datasets in the future.

\begin{figure}
  \centering
  \includegraphics[width=0.95\textwidth]{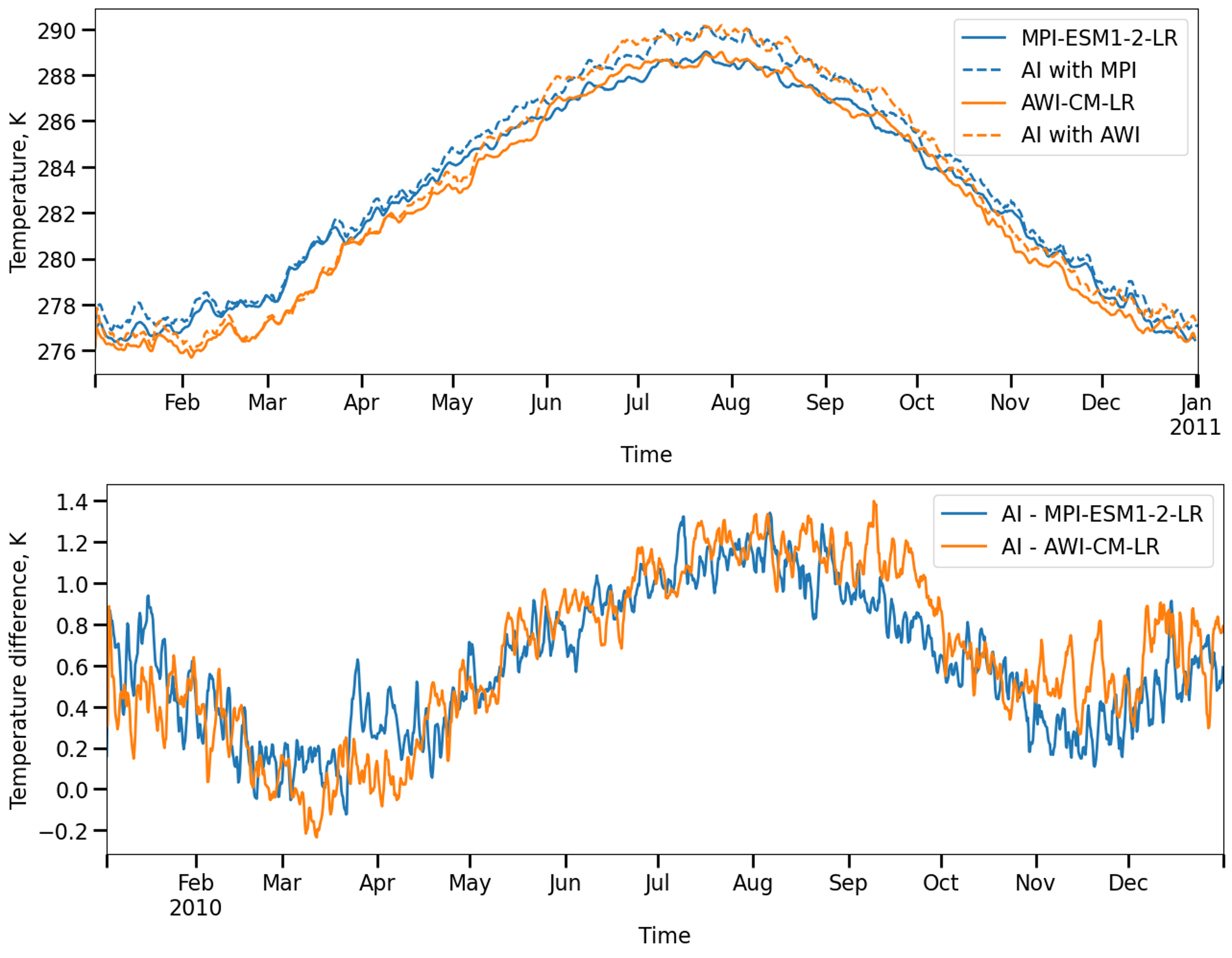}
  \caption{Top: Global mean 2-metre temperature time series for land areas only (6-hour values, 1-day rolling mean) for CMIP6 historical simulations (year 2010) and AI-NWP generated daily forecasts initialized from those simulations. The AI-NWP time series is derived from lead hours 24, 30, 36, and 42 (2-day dataset) of AI-NWP forecasts initiated from CMIP6 model fields. Bottom: Difference between CMIP6 and AI-NWP generated data. Two low-resolution (250km) models are used for initial conditions: AWI-CM-LR and MPI-ESM1-2-LR. The differences between AI-generated data and both models are of similar amplitude.}
  \label{fig:time_series_no_ocean}
\end{figure}

The AI-NWP-based 2-metre temperature time series over land closely follows the variability of climate model data (Figure \ref{fig:time_series_no_ocean}, top). Differences (Figure \ref{fig:time_series_no_ocean}, bottom) are mostly positive (indicating that AI-NWP data are warmer over land), with two pronounced maxima (January, August) and minima (March, November). These differences are remarkably similar between the models. As the considered climate models exhibit mostly positive biases over land compared to ERA5, AI-NWP data over land are closer to realistic temperatures. The range of differences lies between about -0.2\,K and 1.3\,K and falls within the range of CMIP6 model biases. The time series that include ocean data show a negative bias, which is also within the 1\,K range (Figure \ref{fig:A6}).

Initializing AI-NWP with low-resolution climate model data produces forecast fields with improved detail. These fields can be combined into a dataset that compares well with the original data in terms of global mean 2-metre temperature variability and spatial distribution of temperature over land, indicating a correction towards a state closer to ERA5, which is constrained by observations. Although our conclusions are based on two year-long simulations of two low-resolution climate models, the results demonstrate that downscaling of reasonable quality with current AI-NWP is feasible.

\subsection{Initialisation with scenario simulation data}

\begin{figure}
  \centering
  \includegraphics[width=0.95\textwidth]{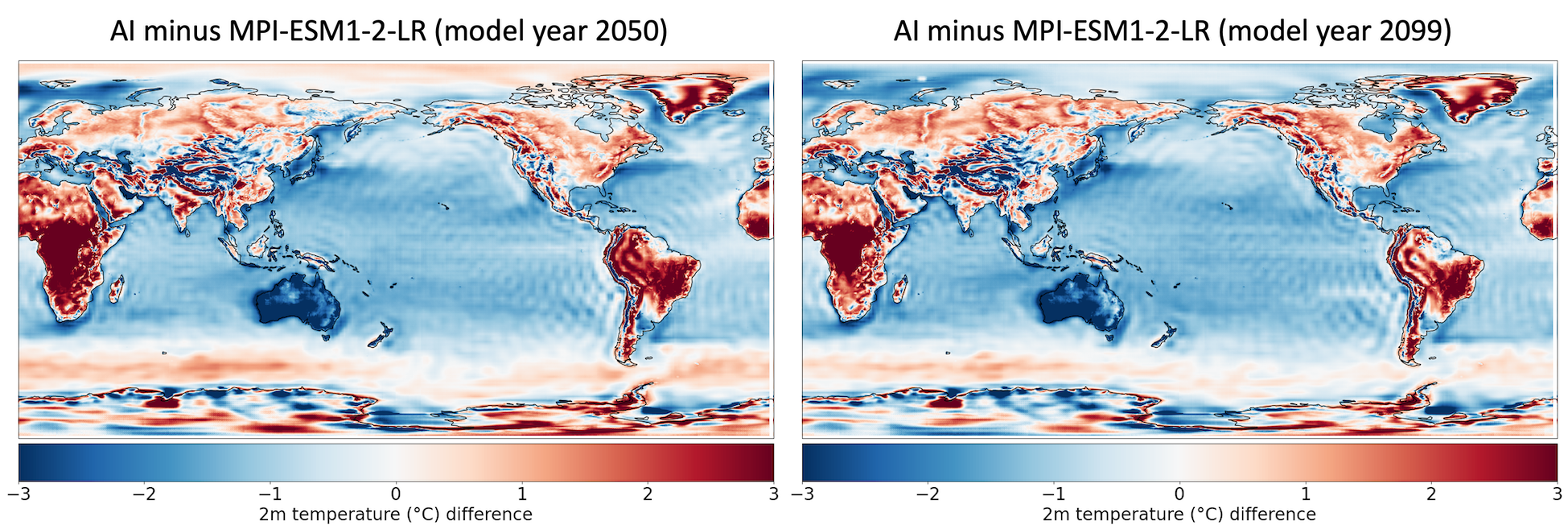}
  \caption{Difference in annual mean 2-metre temperature between AI-NWP generated data and MPI-ESM1-2-LR scenario simulations for model year 2050 (left) and 2099 (right). The AI-NWP time series is derived from lead hours 24, 30, 36, and 42 (2-day dataset) of AI-NWP forecasts initiated from MPI-ESM1-2-LR model fields.}
  \label{fig:fig12}
\end{figure}

To evaluate the applicability of our approach to scenario climate simulations, we generated AI-NWP data for the years 2050 and 2099 using the MPI-ESM2-2-LR model (scenario SPP585). Similarly to the historical simulations, the second forecast days were used. The differences in yearly mean 2-metre temperature between AI-NWP generated data and original climate model data (Figure \ref{fig:fig12}) show patterns and magnitudes similar to those observed in historical simulations (Figure \ref{fig:fig7}). This similarity suggests that the AI-NWP attempts to correct systematic differences related to fast (daily) atmospheric dynamics, such as the generation of low clouds in regions like central Africa and the northern part of South America. Over the oceans, the temperature is generally cooler, with the exception of the Southern Ocean.

\begin{figure}
  \centering
  \includegraphics[width=0.95\textwidth]{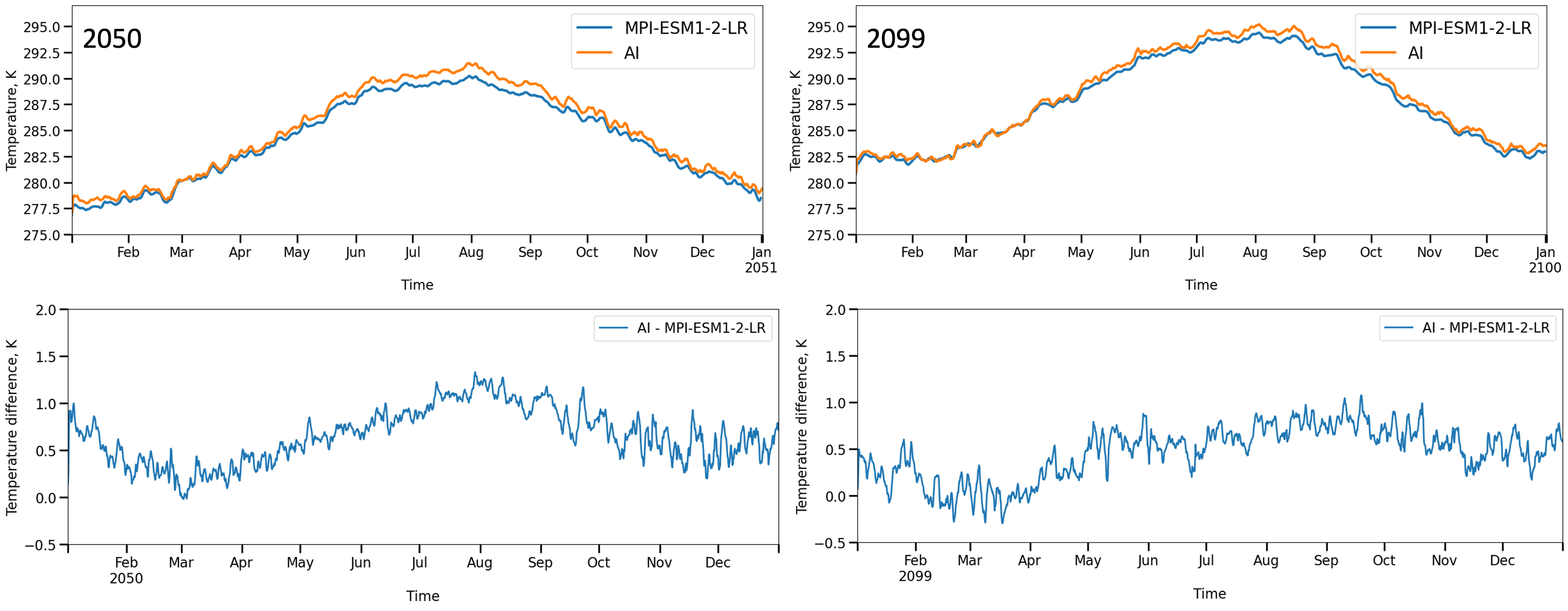}
  \caption{Top: Global mean 2-metre temperature time series for land areas only (6-hour values, 1-day rolling mean) for MPI-ESM1-2-LR scenario simulations (years 2050 and 2099, SSP585) and AI-NWP generated daily forecasts initialized from those simulations. The AI-NWP time series is derived from lead hours 24, 30, 36, and 42 (2-day dataset) of AI-NWP forecasts initiated from CMIP6 model fields. Bottom: Difference between  MPI-ESM1-2-LR scenario simulations and AI-NWP generated data. The difference for both years are the same order of magnitude as for historical simulations.}
  \label{fig:fig11}
\end{figure}

The differences in the time series of 2-metre temperature between AI-NWP generated data and the climate model (Figure \ref{fig:fig11}) for the years 2050 and 2099 are consistent with the results from the historical period (Figure \ref{fig:time_series_no_ocean}). These differences are predominantly positive over land, as shown in Figure \ref{fig:fig11}, and generally fall within the 1\,K range for most of the time. The difference for the year 2099 exhibits a less pronounced seasonal cycle. The discrepancies between AI-based and original data are significantly smaller than the climate change signal.

\begin{figure}
  \centering
  \includegraphics[width=0.95\textwidth]{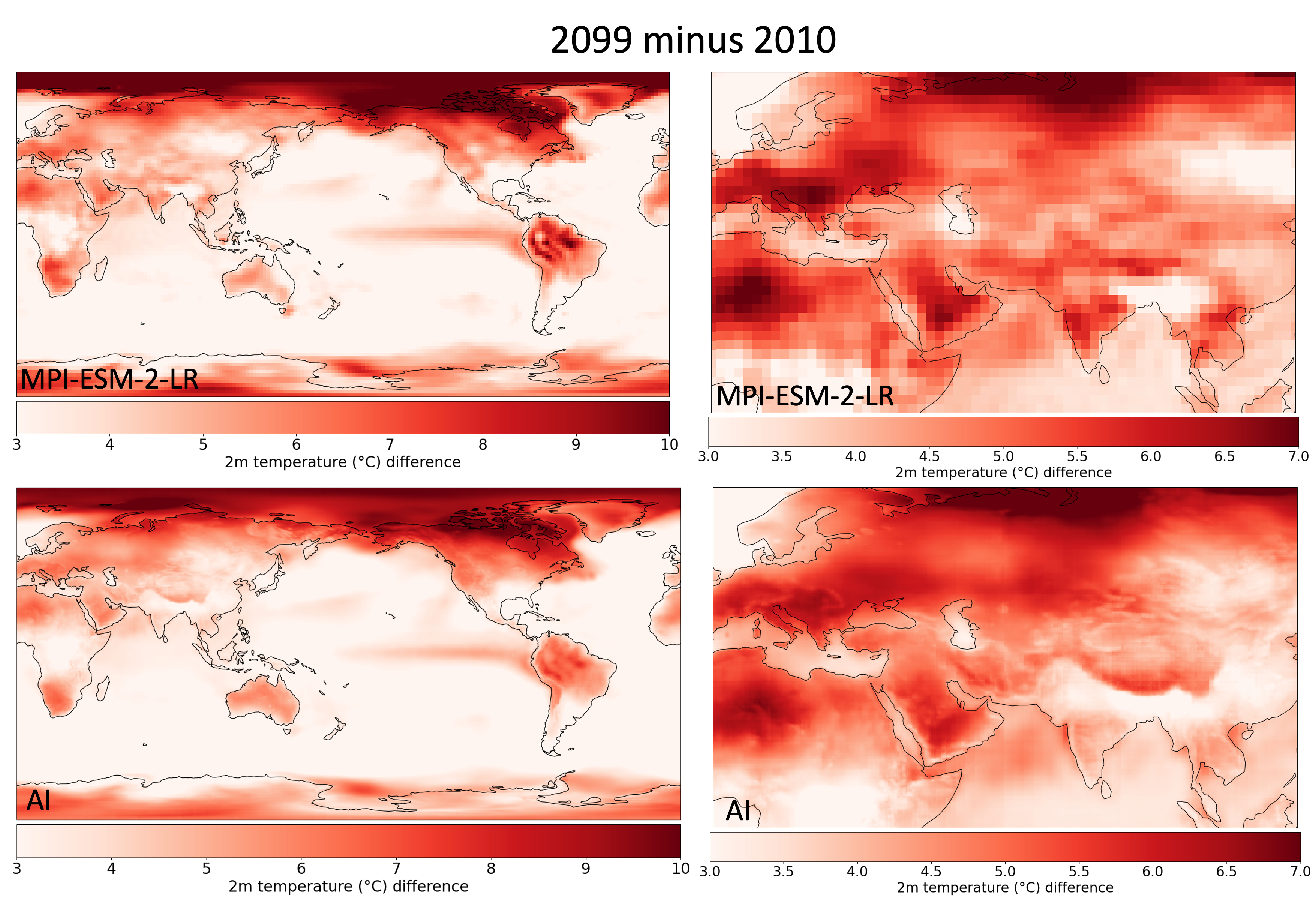}
  \caption{Global mean 2-metre temperature difference between model years of 2010 (historical) and 2099 (scenario, SSP585) of MPI-ESM1-2-LR simulations (top) and AI-NWP data generated from MPI-ESM1-2-LR data. Global view (left) and zoom to regions over Europe, Africa and Asia (right) to demonstrate level of details in the differences. Large scale spatial patterns are mostly the same.}
  \label{fig:fig12}
\end{figure}

To illustrate the potential spatial distribution of the climate change signal following downscaling with AI-NWP, we subtracted the data for the year 2010 from the year 2099 of the MPI-ESM-LR model (Figure \ref{fig:fig12}). For an accurate climate signal, it would be necessary to average over more extended periods from both historical and scenario simulations and account for model drift. However, this figure serves as a representative demonstration. The figure indicates an improved localization of temperature differences, incorporating orographic details. The magnitude of differences between past and future years in the AI-NWP data is smaller and more diffuse compared to the more localized warming hotspots observed in the original MPI-ESM-LR data.

When applied to scenario climate simulations, downscaling with AI-NWP functions similarly to its application to historical simulations. The spatial distribution of 2-metre temperature differences between AI-NWP-based data and climate model data is comparable, as are the differences in global mean 2-metre temperature. This suggests that the differences are more systematic in nature and are not significantly influenced by the application of AI-NWP, which was trained on ERA5 reanalysis, to conditions outside the training envelope.

\section{Discussion and conclusions}

Our study demonstrates that AI-based numerical weather prediction systems (AI-NWP) can effectively downscale low-resolution climate data to generate high-resolution atmospheric fields. By initializing AI-NWP with smoothed ERA5 data and low-resolution CMIP6 climate model data, we found that AI-NWP can develop detailed outputs within one forecast day with a resolution that closely resembles those of the training data. Our approach shows promise for creating high-resolution, long-term datasets from CMIP-type models. Since the AI-NWP models are trained on observationally well-constrained ERA5 reanalysis, they can potentially perform bias correction, moving climate model outputs closer to observed data. More work is, however, required to better understand the distributional shift that the models generate.

The dynamics of atmospheric fields generated by AI-NWP closely resembles that of the original climate models. This is particularly evident in the absence of additional mesoscale dynamics, such as the formation of new cyclones. Unlike regional climate models (RCMs), which can develop such dynamics when applied to sufficiently large regions, our method aligns more closely with statistical downscaling techniques rather than dynamical downscaling. And yet, a significant amount of detail in both cases is triggered by the responses to the details of the underlying surface which were absent in coarse simulations. It is likely that this part of small scale dynamics is rendered correctly in both cases. Moreover the lack of mesoscale dynamics might be a consequence of the limited lead time in our experiments and further work is required to understand if this is an intrinsic limitation.

In this work, we have applied an AI-NWP downscaling method to individual years of climate model simulations. Although we can generate more detailed spatial fields (at higher resolution of $\approx$30\,km), the AI results follow the climate trajectory realised by the coarse-resolution model. Hence, relevant feedbacks from small scale processes to the long-term trajectory of the simulation are still missing or only parameterised, such as feedbacks from mesocale ocean eddies, or the response of clouds to warming. The method presented here can thus not replace kilometre-scale resolution (e.g.\,2--5\,km), eddy- and cloud-resolving climate simulations \citep{hohenegger2023, rackow2024} that potentially resolve very different responses of the coupled atmosphere-ocean system to warming than typical CMIP models.

The resulting high-resolution data from AI-NWP can be used for several applications. They can serve as boundary conditions for further downscaling with regional climate models (RCMs). This eliminates the need for multiple nested downscaling steps, simplifying the process and reducing computational costs. Additionally, if AI-NWP models are enhanced to output hourly or even finer data, they can provide the temporal resolution required by sectors like wind energy for accurate statistical analyses. Beyond boundary conditions, the detailed spatial information generated by AI-NWP is valuable for climate adaptation and mitigation, enabling more precise risk assessments, infrastructure planning, and policy-making at local and regional scales. Aurora, a large-scale foundational model, already produces global output at 0.1$^\circ$ resolution \citep{bodnar2024aurora}, and other AI-NWP models will most probably catch up in the near future.

The fact that initial conditions can be smoothed with only a minor reduction in model skill indicates the potential for distributing lower-resolution initial conditions for initiatives like ECMWF's "forecast in the box", that would allow users to run forecast models locally to recreate full-resolution data. Current 1/4 degree initial conditions contain only few fields, and require just about 140MB of disk space. However, if tens of thousands of users are supposed to be served daily, even small optimizations in the volume of distributed data will count, especially if models use more fields for initial conditions and higher target resolutions. 

This work represents a first step, leveraging AI-NWP tools that are not specifically designed for climate downscaling purposes. A natural next step would be to fine-tune the models for the downscaling task. Alternative approaches, such as foundation models trained on extensive climate datasets and subsequently fine-tuned for global or regional downscaling, may yield better results in the future. However, a compelling aspect of our study is the use of AI-based weather prediction systems, which are readily available and likely to continue receiving substantial resources for development from both scientific and industrial sectors. These systems will likely incorporate the most advanced atmospheric data for training, optimized for various hardware configurations and have user-friendly interfaces. Consequently, directly applying AI-NWP to climate data could prove more straightforward and lead to better skill than developing specialized solutions. Moreover, our approach offers potential for identifying deficiencies in AI-NWP models, such as the too cold temperature over the ocean for Pangu-Weather, and incorporates a built-in bias correction mechanism. 

Future research could focus on testing other AI-NWP models and expanding the analysis to include a wider range of variables. We plan to downscale complete historical and scenario simulations for various climate models, particularly those representing different ends of the spectrum in terms of climate sensitivity and the quality of historical simulations. This will help refine the methodology, enhance the accuracy of downscaled data, and help to understand the applicability of AI-NWP in diverse climatic contexts, in particular if ERA5-trained models can be used for different scenario runs.  It will also be  interesting to compare results of the downscaling with RCMs and with AI-NWP, as well as to try driving RCMs with AI-NWP generated data.

\section{Acknowlegements}
The computing time and storage for Pangu-Weather simulations and data processing is provided by Deutsches Klimarechenzentrum (DKRZ). 
Nikolay Koldunov and Sergey Danilov are supported by projects S1: Diagnosis and Metrics in Climate Models and M3: Towards Consistent Subgrid Momentum Closures of the Collaborative Research Centre TRR 181 “Energy Transfer in Atmosphere and Ocean”, funded by the Deutsche Forschungsgemeinschaft (DFG, German Research Foundation, project no. 274762653). Thomas Rackow has been supported by the European Commission Horizon 2020 Framework Programme nextGEMS (grant no. 101003470). Suvarchal K. Cheedela is supported by BMBF-funded project WarmWorld. Thomas Jung is supported by the EERIE project (Grant Agreement No 101081383) funded by the European Union. Views and opinions expressed are however those of the authors only and do not necessarily reflect those of the European Union or the European Climate Infrastructure and Environment Executive Agency (CINEA). Neither the European Union nor the granting authority can be held responsible for them.

\bibliographystyle{apalike}

\appendix

\renewcommand{\thefigure}{\thesection\arabic{figure}}
\setcounter{figure}{0}

\section{Appendix A: Additional Information}
This appendix contains supplementary figures.

\clearpage

\begin{figure}
  \centering
  \includegraphics[width=0.9\textwidth]{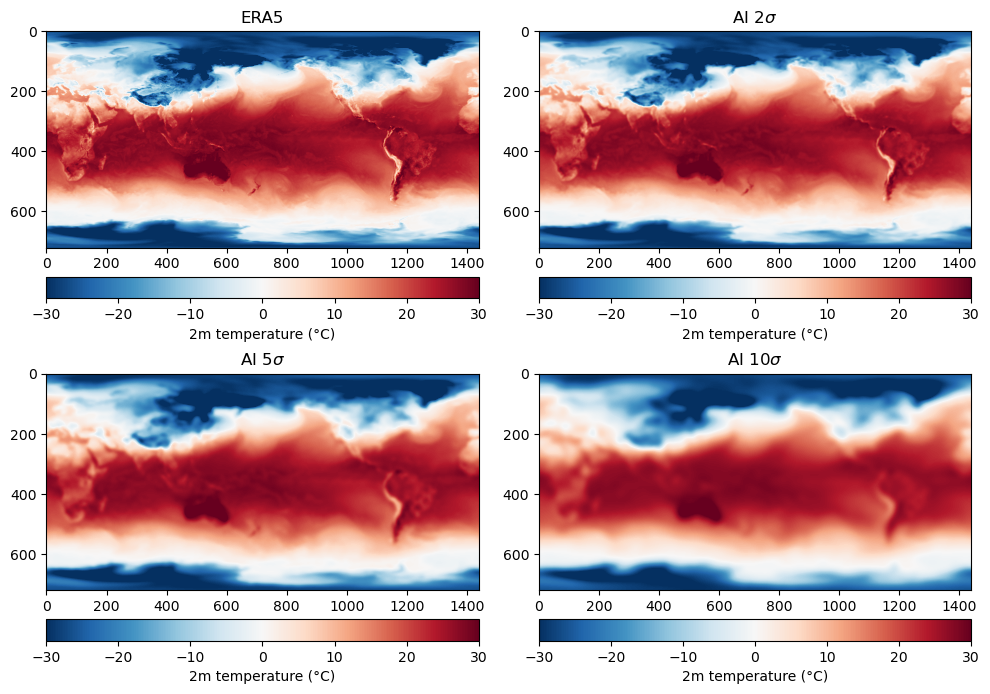}
  \caption{The 2-metre temperature used as part of the initial conditions (2020-01-01T00:00:00) with several levels of Gaussian smoothing. All other initial fields were smoothed similarly to emulate the fields generated by atmospheric models with different horizontal resolutions. Global view.}
  \label{fig:A1}
\end{figure}

\begin{figure}
  \centering
  \includegraphics[width=0.9\textwidth]{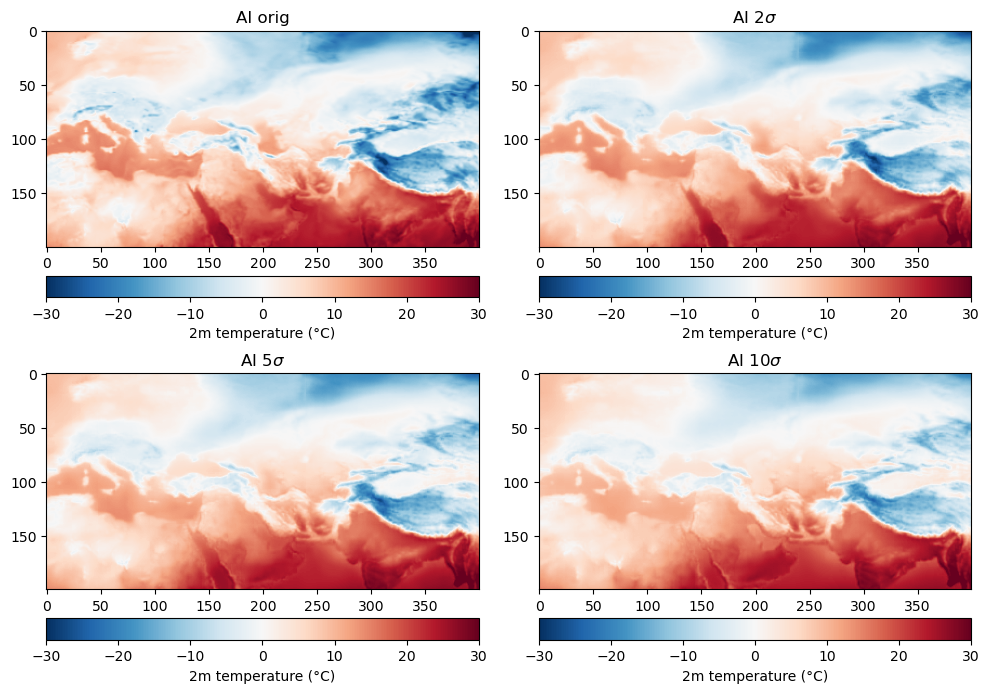}
  \caption{
  The 2-metre temperature after 6 hours of forecast (2020-01-01T06:00:00) using the AI-WPS for simulations with varying smoothing of initial conditions. Already after 6 hours the AI-WPS shows good level of detail comparable to the fields it was originally trained on (ERA5).}
  \label{fig:A2}
\end{figure}

\begin{figure}
  \centering
  \includegraphics[width=0.95\textwidth]{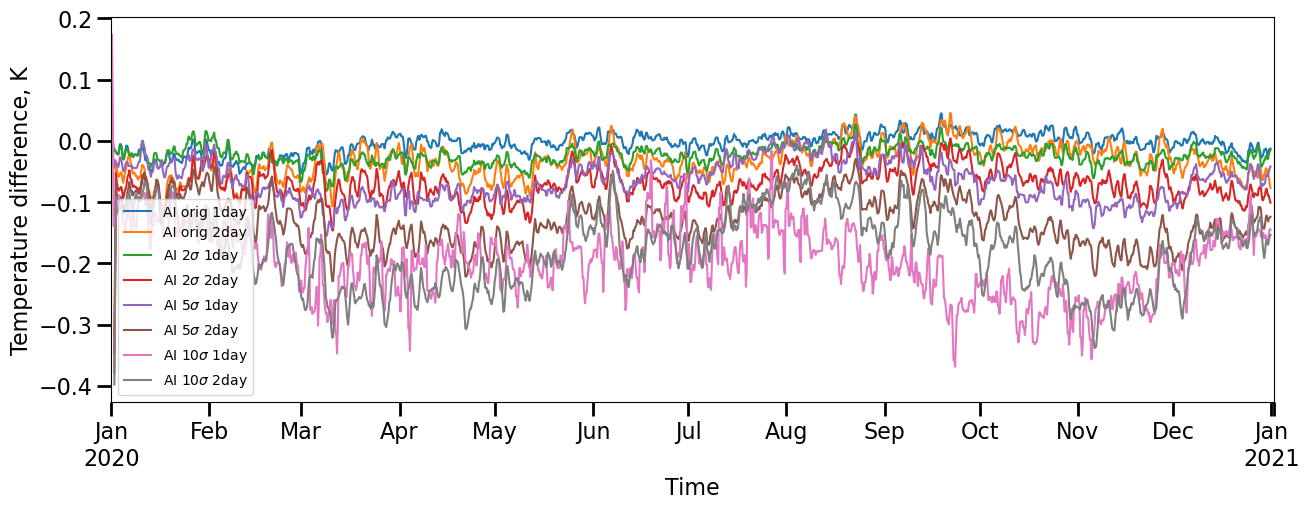}
  \caption{Difference in global mean 2-metre temperature time series, constructed from AI-NWP data with varying initial condition smoothing and lead times (1-day, 2-day), compared to ERA5 time series (6-hour values, daily rolling mean). These differences correspond to the graphs in Fig. \ref{fig:fig3}. Mean values are listed in Table \ref{table:smoothing}.}
  \label{fig:A3}
\end{figure}

\begin{figure}
  \centering
  \includegraphics[width=0.95\textwidth]{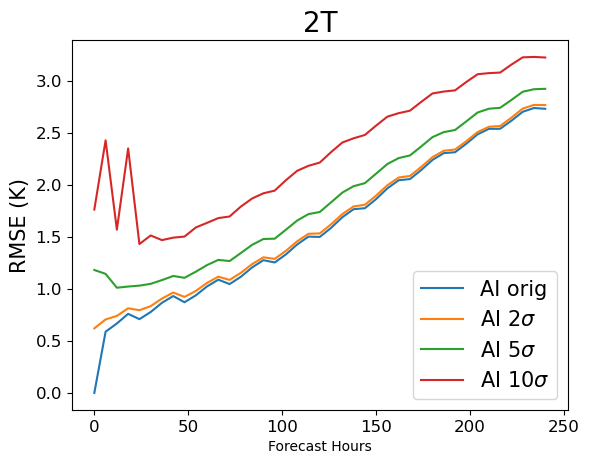}
  \caption{Root Mean Square Error (RMSE) of global 2-metre temperature at different levels of smoothing relative to ERA5. Mean values over one year (2020).}
  \label{fig:A4}
\end{figure}

\begin{figure}
  \centering
  \includegraphics[width=0.95\textwidth]{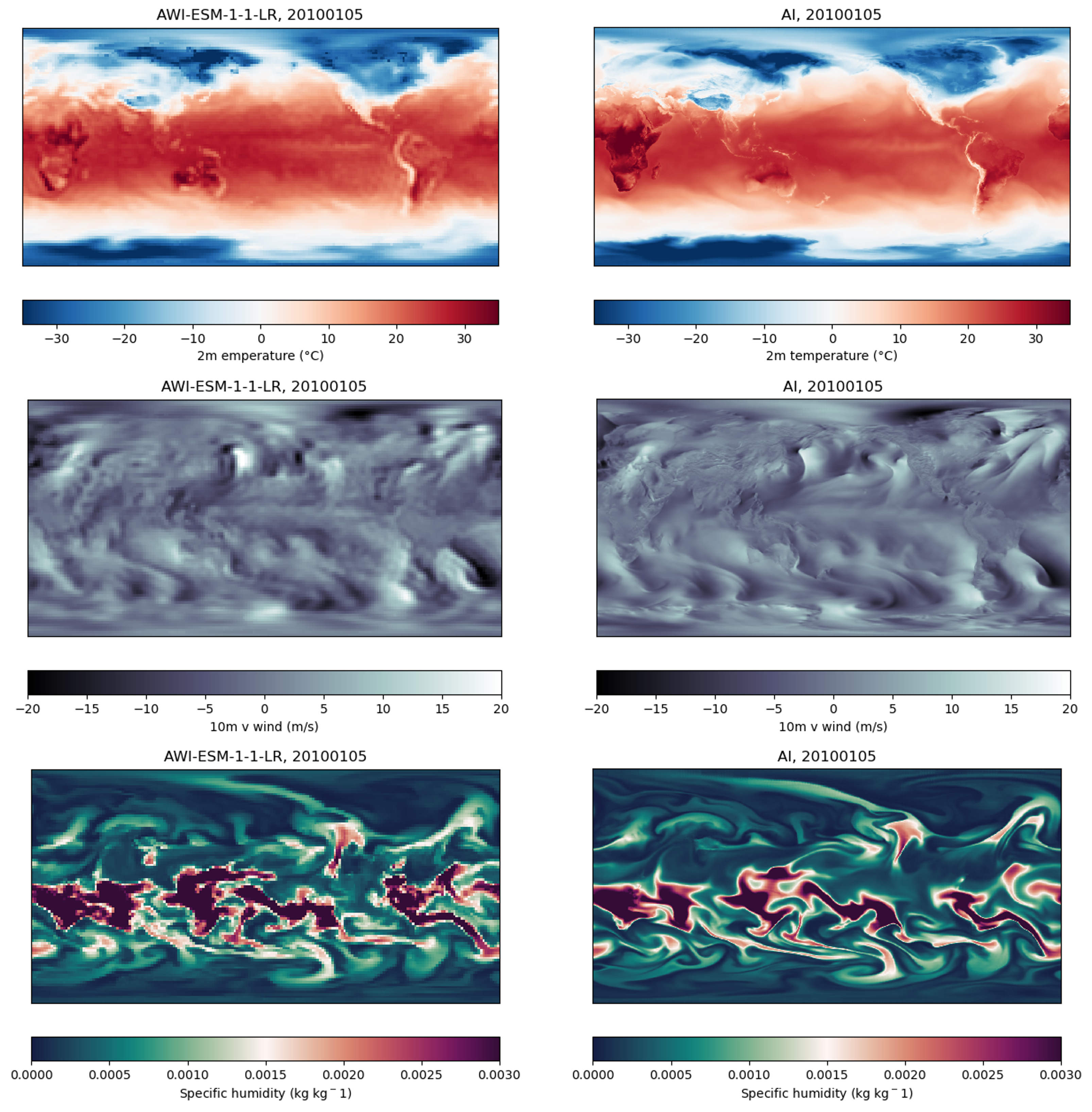}
  \caption{Snapshot from 2010-01-05 comparing original fields from a low-resolution climate model (left) and the result of an AI-NWP forecast initialized 2 days prior from climate model data (right). Same as Fig \ref{fig:fig6}, but global.}
  \label{fig:A5}
\end{figure}

\begin{figure}
  \centering
  \includegraphics[width=0.95\textwidth]{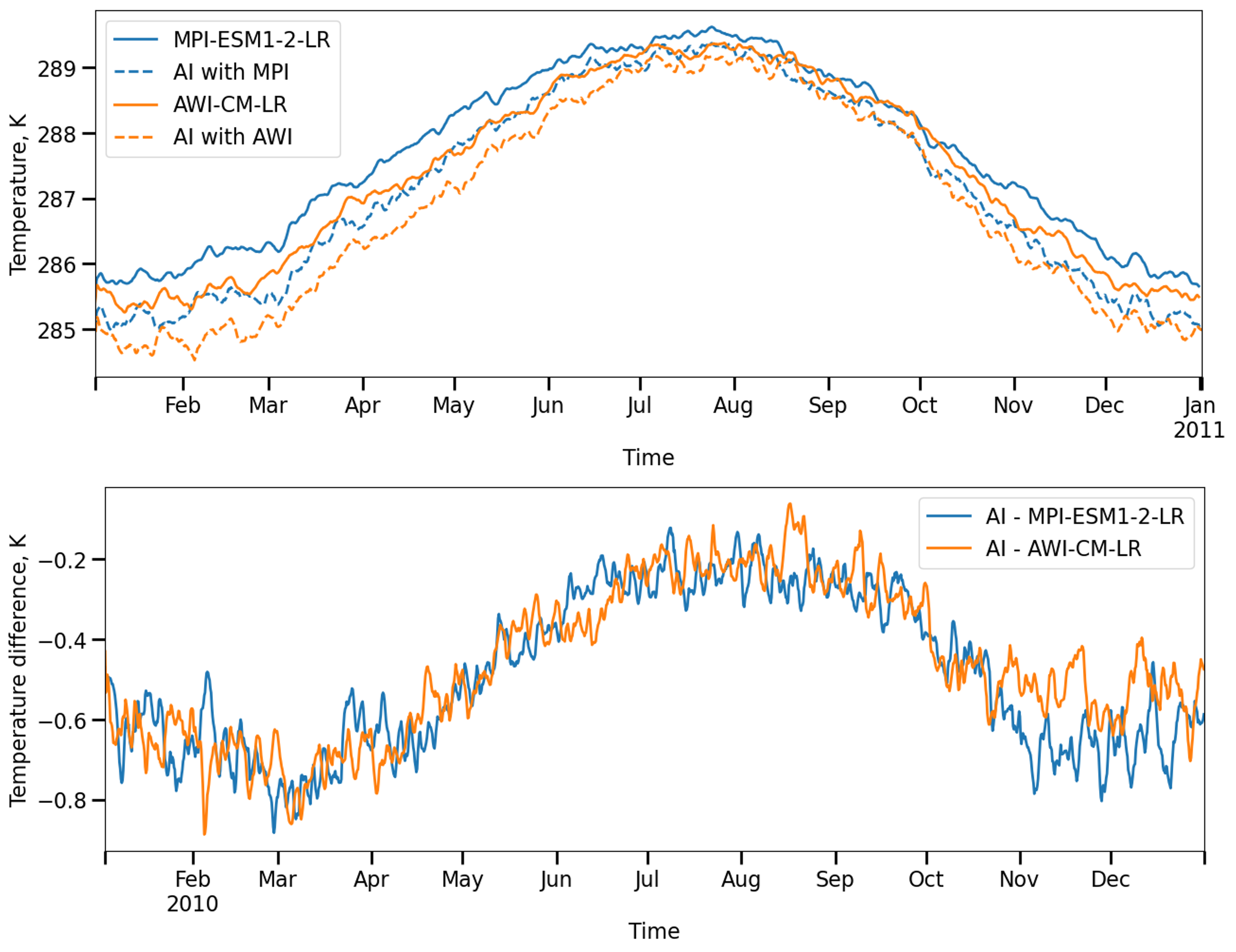}
  \caption{Top: Global mean 2-metre temperature time series (6-hour values, 1-day rolling mean) for CMIP6 historical simulations (year 2010) and AI-NWP generated daily forecasts initialized from those simulations. The AI-NWP time series is derived from lead hours 24, 30, 36, and 42 (2-day dataset) of AI-NWP forecasts initiated from CMIP6 model fields. Bottom: Difference between CMIP6 and AI-NWP generated data. Two low-resolution (250km) models are used: AWI-CM-LR and MPI-ESM1-2-LR. Both models share the same ECHAM atmosphere but use different ocean components. The differences between AI-generated data and both models are of similar amplitude.}
  \label{fig:A6}
\end{figure}

\begin{figure}
  \centering
  \includegraphics[width=0.95\textwidth]{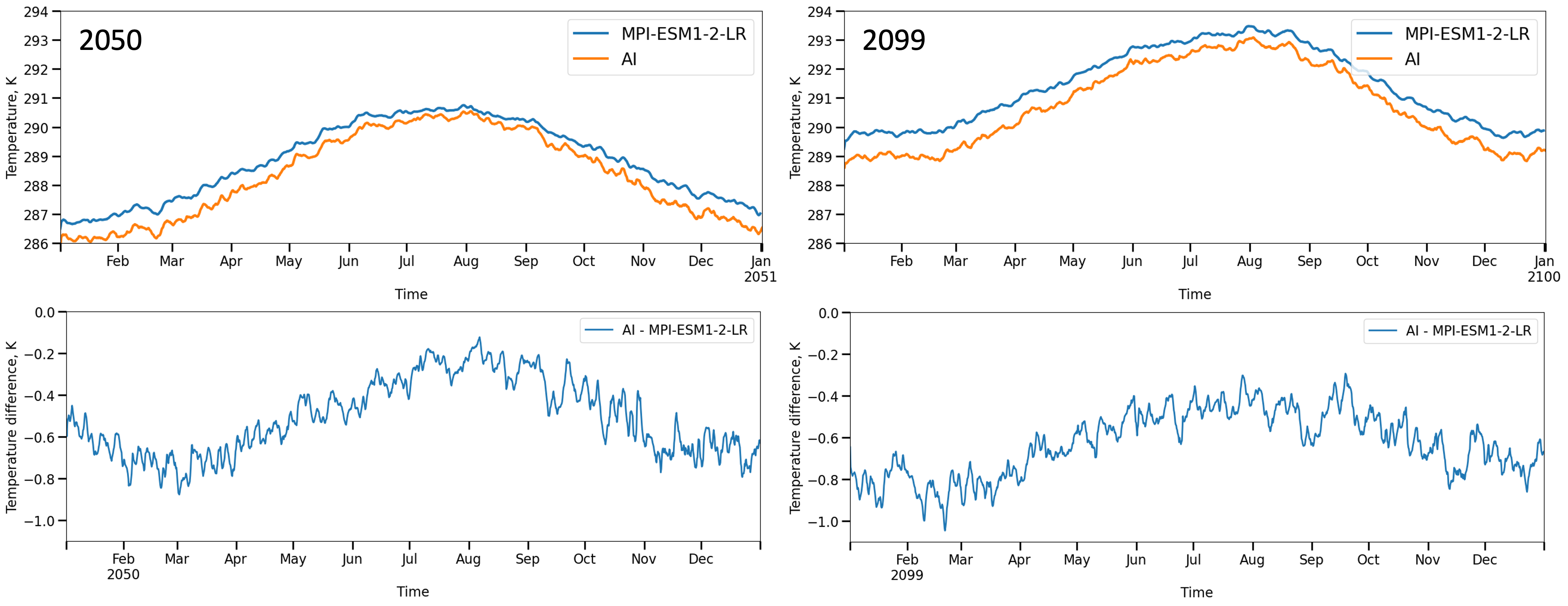}
  \caption{Top: Global mean 2-metre temperature time series (6-hour values, 1-day rolling mean) for MPI-ESM1-2-LR scenario simulations (years 2050 and 2099, SSP585) and AI-NWP generated daily forecasts initialized from those simulations. The AI-NWP time series is derived from lead hours 24, 30, 36, and 42 (2-day dataset) of AI-NWP forecasts initiated from CMIP6 model fields. Bottom: Difference between  MPI-ESM1-2-LR scenario simulations and AI-NWP generated data.}
  \label{fig:fig10}
\end{figure}

\clearpage
\renewcommand{\thefigure}{\thesection\arabic{figure}}
\setcounter{figure}{0}
\section{Appendix B: Data locality use cases}
We aim to demonstrate the behavior of downscaled climate variables by examining their time series around areas exhibiting potentially strong spatial gradients. This demonstration will underscore the advantages of creating high-resolution versions of low-resolution climate model datasets for adaptation and mitigation applications.

\begin{figure}
  \centering
  \includegraphics[width=0.90\textwidth]{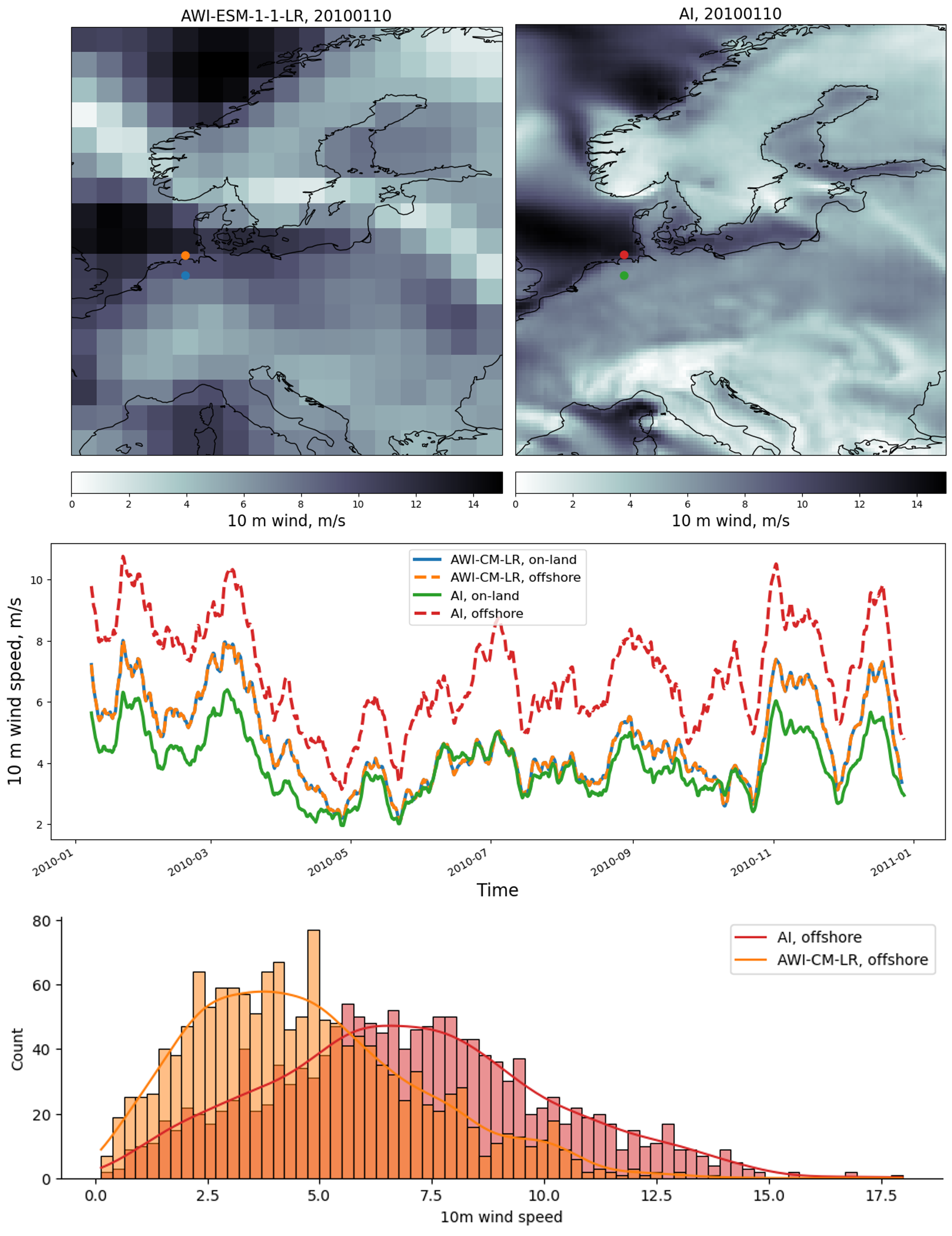}
  \caption{Top: Wind speed at 10m from AWI-ESM-LR (left) and AI-NWP generated data (right). The analyzed points are indicated. Data for the blue and orange points are extracted from the low-resolution AWI-ESM-LR data, falling into the same grid point, despite being on-land and offshore respectively. Data for the green and red points are extracted from AI-NWP generated data, falling into different grid points. Middle: Time series of 10m wind speed for the indicated points. The blue and dashed orange lines coincide exactly, as the points fall within the same grid box. A clear difference is observed between on-land and offshore points in the AI-generated data. Bottom: Histogram of 10m wind speeds for offshore points in AWI-ESM-LR and AI-NWP generated data. The downscaled information provides users with enhanced locality}
  \label{fig:B1}
\end{figure}

First, we analyze the wind speed data (Figure \ref{fig:B1}). Two points were selected, one on-land and the other offshore. Data were extracted from the grid points nearest to the selected locations for both a low-resolution climate model and the corresponding AI-NWP dataset. The selection ensured that, for the low-resolution model, both points fell within a single grid point, although the on-land grid point was situated several hundred kilometers inland.

The time series of wind speed plotted for the low-resolution climate model are naturally identical for both points, while they show significant differences in the AI-NWP data. As anticipated, high-resolution data indicate stronger winds offshore, with differences reaching several meters per second (Figure \ref{fig:B1}, middle row). Similarly, the on-land point shows lower wind speeds on average. High-resolution AI-NWP data demonstrate more frequent occurrences of stronger winds, reaching higher speeds (Figure \ref{fig:B1}, bottom row).

Even small differences in wind speed can lead to substantial differences in wind energy production, given that wind power increases approximately with the cube of the wind speed. Therefore, more accurate wind speed estimates are crucial for adaptation use cases related to wind energy.

\begin{figure}
  \centering
  \includegraphics[width=0.90\textwidth]{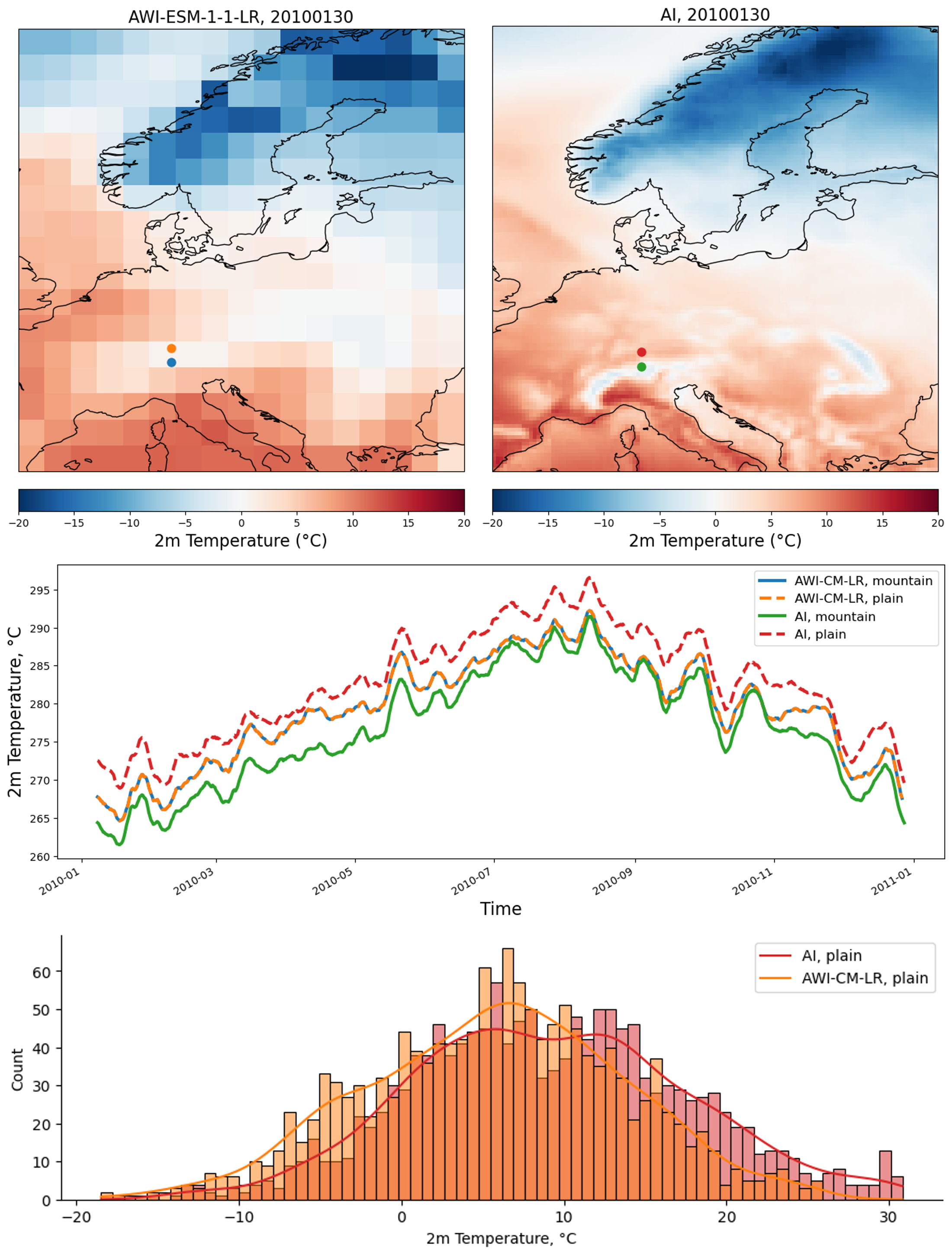}
  \caption{Same ad Fig \ref{fig:B1}, but for temperature, and points located on over the mountains, and on the plain.}
  \label{fig:B2}
\end{figure}

Another regions with strong spatial gradients are mountainous areas. Similarly to the wind use case, we select points that in reality are located in the Alps and on flatter area next to them (Figure \ref{fig:B2}, top row). For low resolution data the nearest grid point is the same, while in AI-NWP generated data there is few degree difference between  mountainous and plain areas (Figure \ref{fig:B2}, middle row). Histogram of temperature values shows that warmer temperatures occurs more often in AI-NWP generated data than anticipated by the low resolution climate model (Figure \ref{fig:B2}, bottom row).

\begin{thebibliography}{}

\bibitem[Ba{\~n}o-Medina et~al., 2020]{bano2020configuration}
Ba{\~n}o-Medina, J., Manzanas, R., and Guti{\'e}rrez, J.~M. (2020).
\newblock Configuration and intercomparison of deep learning neural models for
  statistical downscaling.
\newblock {\em Geoscientific Model Development}, 13(4):2109--2124.

\bibitem[Ben-Bouallegue et~al., 2023]{BenBouallegue2023}
Ben-Bouallegue, Z., Clare, M. C.~A., Magnusson, L., Gascon, E., Maier-Gerber,
  M., Janousek, M., Rodwell, M., Pinault, F., Dramsch, J.~S., Lang, S. T.~K.,
  Raoult, B., Rabier, F., Chevallier, M., Sandu, I., Dueben, P., Chantry, M.,
  and Pappenberger, F. (2023).
\newblock The rise of data-driven weather forecasting.

\bibitem[Bi et~al., 2023]{bi2023accurate}
Bi, K., Xie, L., Zhang, H., Chen, X., Gu, X., and Tian, Q. (2023).
\newblock Accurate medium-range global weather forecasting with {3D} neural
  networks.
\newblock {\em Nature}, 619(7970):533--538.

\bibitem[Bodnar et~al., 2024]{bodnar2024aurora}
Bodnar, C., Bruinsma, W.~P., Lucic, A., Stanley, M., Brandstetter, J., Garvan,
  P., Riechert, M., Weyn, J., Dong, H., Vaughan, A., Gupta, J.~K., Tambiratnam,
  K., Archibald, A., Heider, E., Welling, M., Turner, R.~E., and Perdikaris, P.
  (2024).
\newblock Aurora: A foundation model of the atmosphere.

\bibitem[Chen et~al., 2023]{chen2023fuxi}
Chen, L., Zhong, X., Zhang, F., Cheng, Y., Xu, Y., Qi, Y., and Li, H. (2023).
\newblock {F}u{X}i: a cascade machine learning forecasting system for 15-day
  global weather forecast.
\newblock {\em npj Climate and Atmospheric Science}, 6(1):190.

\bibitem[Damm et~al., 2020]{damm2020market}
Damm, A., K{\"o}berl, J., Stegmaier, P., Alonso, E.~J., and Harjanne, A.
  (2020).
\newblock The market for climate services in the tourism sector--{A}n analysis
  of {A}ustrian stakeholders’ perceptions.
\newblock {\em Climate Services}, 17:100094.

\bibitem[Danek et~al., 2020]{awiesmdata}
Danek, C., Shi, X., Stepanek, C., Yang, H., Barbi, D., Hegewald, J., and
  Lohmann, G. (2020).
\newblock {AWI AWI-ESM1.1LR} model output prepared for {CMIP6 CMIP} historical.

\bibitem[ECMWF, 2024]{climetlab}
ECMWF (2024).
\newblock climetlab, https://github.com/ecmwf/climetlab.
\newblock Accessed: 2024-06-23.

\bibitem[Ekstr{\"o}m et~al., 2015]{ekstrom2015stat}
Ekstr{\"o}m, M., Grose, M.~R., and Whetton, P.~H. (2015).
\newblock An appraisal of downscaling methods used in climate change research.
\newblock {\em Wiley Interdisciplinary Reviews: Climate Change}, 6(3):301--319.

\bibitem[Fan et~al., 2020]{fan2020global}
Fan, X., Duan, Q., Shen, C., Wu, Y., and Xing, C. (2020).
\newblock {Global surface air temperatures in CMIP6: Historical performance and
  future changes}.
\newblock {\em Environmental Research Letters}, 15(10):104056.

\bibitem[Fowler et~al., 2007]{fowler2007stat}
Fowler, H.~J., Blenkinsop, S., and Tebaldi, C. (2007).
\newblock Linking climate c hange modelling to impacts studies.

\bibitem[Giorgetta et~al., 2013]{giorgetta2013climate}
Giorgetta, M.~A., Jungclaus, J., Reick, C.~H., Legutke, S., Bader, J.,
  B{\"o}ttinger, M., Brovkin, V., Crueger, T., Esch, M., Fieg, K., et~al.
  (2013).
\newblock Climate and carbon cycle changes from 1850 to 2100 in {MPI-ESM}
  simulations for the {C}oupled {M}odel {I}ntercomparison {P}roject phase 5.
\newblock {\em Journal of Advances in Modeling Earth Systems}, 5(3):572--597.

\bibitem[Giorgi, 2019]{giorgi2019rcm}
Giorgi, F. (2019).
\newblock Thirty years of regional climate modeling: where are we and where are
  we going next?
\newblock {\em Journal of Geophysical Research: Atmospheres},
  124(11):5696--5723.

\bibitem[Giorgi and Gutowski~Jr, 2015]{giorgi2015cordex}
Giorgi, F. and Gutowski~Jr, W.~J. (2015).
\newblock Regional dynamical downscaling and the {CORDEX} initiative.
\newblock {\em Annual review of environment and resources}, 40:467--490.

\bibitem[Haarsma et~al., 2016]{haarsma2016high}
Haarsma, R.~J., Roberts, M.~J., Vidale, P.~L., Senior, C.~A., Bellucci, A.,
  Bao, Q., Chang, P., Corti, S., Fu{\v{c}}kar, N.~S., Guemas, V., et~al.
  (2016).
\newblock {High resolution model intercomparison project (HighResMIP v1. 0) for
  CMIP6}.
\newblock {\em Geoscientific Model Development}, 9(11):4185--4208.

\bibitem[Hakim and Masanam, 2023]{Hakim2023}
Hakim, G.~J. and Masanam, S. (2023).
\newblock Dynamical tests of a deep-learning weather prediction model.

\bibitem[Hanke et~al., 2016]{hanke2016yac}
Hanke, M., Redler, R., Holfeld, T., and Yastremsky, M. (2016).
\newblock {YAC} 1.2. 0: new aspects for coupling software in {E}arth system
  modelling.
\newblock {\em Geoscientific Model Development}, 9(8):2755--2769.

\bibitem[Harris et~al., 2022]{harris2022generative}
Harris, L., McRae, A.~T., Chantry, M., Dueben, P.~D., and Palmer, T.~N. (2022).
\newblock A generative deep learning approach to stochastic downscaling of
  precipitation forecasts.
\newblock {\em Journal of Advances in Modeling Earth Systems},
  14(10):e2022MS003120.

\bibitem[Hauser and Developers, 2024]{regionmask}
Hauser, M. and Developers (2024).
\newblock regionmask, https://github.com/regionmask/regionmask/.
\newblock Accessed: 2024-06-23.

\bibitem[Hersbach et~al., 2020]{hersbach2020era5}
Hersbach, H., Bell, B., Berrisford, P., Hirahara, S., Hor{\'a}nyi, A.,
  Mu{\~n}oz-Sabater, J., Nicolas, J., Peubey, C., Radu, R., Schepers, D.,
  et~al. (2020).
\newblock The era5 global reanalysis.
\newblock {\em Quarterly Journal of the Royal Meteorological Society},
  146(730):1999--2049.

\bibitem[Hohenegger et~al., 2023]{hohenegger2023}
Hohenegger, C., Korn, P., Linardakis, L., Redler, R., Schnur, R., Adamidis, P.,
  Bao, J., Bastin, S., Behravesh, M., Bergemann, M., Biercamp, J., Bockelmann,
  H., Brokopf, R., Br\"uggemann, N., Casaroli, L., Chegini, F., Datseris, G.,
  Esch, M., George, G., Giorgetta, M., Gutjahr, O., Haak, H., Hanke, M.,
  Ilyina, T., Jahns, T., Jungclaus, J., Kern, M., Klocke, D., Kluft, L.,
  K\"olling, T., Kornblueh, L., Kosukhin, S., Kroll, C., Lee, J., Mauritsen,
  T., Mehlmann, C., Mieslinger, T., Naumann, A.~K., Paccini, L., Peinado, A.,
  Praturi, D.~S., Putrasahan, D., Rast, S., Riddick, T., Roeber, N., Schmidt,
  H., Schulzweida, U., Sch\"utte, F., Segura, H., Shevchenko, R., Singh, V.,
  Specht, M., Stephan, C.~C., von Storch, J.-S., Vogel, R., Wengel, C.,
  Winkler, M., Ziemen, F., Marotzke, J., and Stevens, B. (2023).
\newblock {ICON-Sapphire}: simulating the components of the {Earth} system and
  their interactions at kilometer and subkilometer scales.
\newblock {\em Geoscientific Model Development}, 16(2):779--811.

\bibitem[H{\"o}hlein et~al., 2020]{hohlein2020comparative}
H{\"o}hlein, K., Kern, M., Hewson, T., and Westermann, R. (2020).
\newblock A comparative study of convolutional neural network models for wind
  field downscaling.
\newblock {\em Meteorological Applications}, 27(6):e1961.

\bibitem[Hoyer and Hamman, 2017]{hoyer2017xarray}
Hoyer, S. and Hamman, J. (2017).
\newblock xarray: {N-D} labeled arrays and datasets in {Python}.
\newblock {\em Journal of Open Research Software}, 5(1).

\bibitem[Hunter, 2007]{matplotlib}
Hunter, J.~D. (2007).
\newblock {Matplotlib: A 2D graphics environment}.
\newblock {\em Computing in Science \& Engineering}, 9(3):90--95.

\bibitem[Jungclaus et~al., 2013]{jungclaus2013characteristics}
Jungclaus, J.~H., Fischer, N., Haak, H., Lohmann, K., Marotzke, J., Matei, D.,
  Mikolajewicz, U., Notz, D., and Von~Storch, J. (2013).
\newblock {Characteristics of the ocean simulations in the Max Planck Institute
  Ocean Model (MPIOM) the ocean component of the MPI-Earth system model}.
\newblock {\em Journal of Advances in Modeling Earth Systems}, 5(2):422--446.

\bibitem[Koldunov et~al., 2016]{koldunov2016identifying}
Koldunov, N.~V., Kumar, P., Rasmussen, R., Ramanathan, A., Nesje, A.,
  Engelhardt, M., Tewari, M., Haensler, A., and Jacob, D. (2016).
\newblock Identifying climate change information needs for the himalayan
  region: Results from the {GLACINDIA} stakeholder workshop and training
  program.
\newblock {\em Bulletin of the American Meteorological Society},
  97(2):ES37--ES40.

\bibitem[Lam et~al., 2023]{lam2023learning}
Lam, R., Sanchez-Gonzalez, A., Willson, M., Wirnsberger, P., Fortunato, M.,
  Alet, F., Ravuri, S., Ewalds, T., Eaton-Rosen, Z., Hu, W., et~al. (2023).
\newblock Learning skillful medium-range global weather forecasting.
\newblock {\em Science}, 382(6677):1416--1421.

\bibitem[Lang et~al., 2024]{lang2024aifs}
Lang, S., Alexe, M., Chantry, M., Dramsch, J., Pinault, F., Raoult, B., Clare,
  M.~C., Lessig, C., Maier-Gerber, M., Magnusson, L., et~al. (2024).
\newblock Aifs-ecmwf's data-driven forecasting system.
\newblock {\em arXiv preprint arXiv:2406.01465}.

\bibitem[Lessig et~al., 2023]{lessig2023atmorep}
Lessig, C., Luise, I., Gong, B., Langguth, M., Stadtler, S., and Schultz, M.
  (2023).
\newblock {AtmoRep}: A stochastic model of atmosphere dynamics using large
  scale representation learning.

\bibitem[Mardani et~al., 2023]{mardani2023residual}
Mardani, M., Brenowitz, N., Cohen, Y., Pathak, J., Chen, C.-Y., Liu, C.-C.,
  Vahdat, A., Kashinath, K., Kautz, J., and Pritchard, M. (2023).
\newblock Residual diffusion modeling for km-scale atmospheric downscaling.

\bibitem[{Met Office}, 2015]{cartopy}
{Met Office} (2010 - 2015).
\newblock {\em {Cartopy: a cartographic python library with a Matplotlib
  interface}}.
\newblock Exeter, Devon.

\bibitem[Miles et~al., 2024]{zarr}
Miles, A., jakirkham, Bussonnier, M., Moore, J., Orfanos, D.~P., Bennett, D.,
  Stansby, D., Hamman, J., Bourbeau, J., Fulton, A., Lee, G., Abernathey, R.,
  Rzepka, N., Patel, Z., Kristensen, M. R.~B., Verma, S., Chopra, S., Rocklin,
  M., AWA, A.~B., Jones, M., Durant, M., de~Andrade, E.~S., Schut, V., raphael
  dussin, Chaudhary, S., Barnes, C., Nunez-Iglesias, J., and shikharsg (2024).
\newblock zarr-developers/zarr-python: v3.0.0-alpha.

\bibitem[Nguyen et~al., 2023]{nguyen2023climax}
Nguyen, T., Brandstetter, J., Kapoor, A., Gupta, J.~K., and Grover, A. (2023).
\newblock {ClimaX}: A foundation model for weather and climate.

\bibitem[Pathak et~al., 2022]{pathak2022fourcastnet}
Pathak, J., Subramanian, S., Harrington, P., Raja, S., Chattopadhyay, A.,
  Mardani, M., Kurth, T., Hall, D., Li, Z., Azizzadenesheli, K., et~al. (2022).
\newblock Fourcastnet: A global data-driven high-resolution weather model using
  adaptive fourier neural operators.
\newblock {\em arXiv preprint arXiv:2202.11214}.

\bibitem[Rackow et~al., 2018]{rackow2018awicm}
Rackow, T., Goessling, H.~F., Jung, T., Sidorenko, D., Semmler, T., Barbi, D.,
  and Handorf, D. (2018).
\newblock Towards multi-resolution global climate modeling with {ECHAM6-FESOM}.
  {Part II}: climate variability.
\newblock {\em Climate Dynamics}, 50:2369--2394.

\bibitem[Rackow et~al., 2024]{rackow2024}
Rackow, T., Pedruzo-Bagazgoitia, X., Becker, T., Milinski, S., Sandu, I.,
  Aguridan, R., Bechtold, P., Beyer, S., Bidlot, J., Boussetta, S.,
  Diamantakis, M., Dueben, P., Dutra, E., Forbes, R., Goessling, H.~F., Hadade,
  I., Hegewald, J., Keeley, S., Kluft, L., Koldunov, N., Koldunov, A.,
  K\"olling, T., Kousal, J., Mogensen, K., Quintino, T., Polichtchouk, I.,
  S\'arm\'any, D., Sidorenko, D., Streffing, J., S\"utzl, B., Takasuka, D.,
  Tietsche, S., Valentini, M., Vanni\`ere, B., Wedi, N., Zampieri, L., and
  Ziemen, F. (2024).
\newblock Multi-year simulations at kilometre scale with the {Integrated
  Forecasting System} coupled to {FESOM2.5/NEMOv3.4}.
\newblock {\em EGUsphere}, 2024:1--59.

\bibitem[Rampal et~al., 2022]{rampal2022high}
Rampal, N., Gibson, P.~B., Sood, A., Stuart, S., Fauchereau, N.~C., Brandolino,
  C., Noll, B., and Meyers, T. (2022).
\newblock High-resolution downscaling with interpretable deep learning:
  Rainfall extremes over {N}ew {Z}ealand.
\newblock {\em Weather and Climate Extremes}, 38:100525.

\bibitem[Ranasinghe et~al., 2021]{ranasinghe2021ipcc}
Ranasinghe, R., Ruane, A.~C., Vautard, R., Arnell, N., Coppola, E., Cruz,
  F.~A., Dessai, S., Saiful~Islam, A., Rahimi, M., Carrascal, D.~R., et~al.
  (2021).
\newblock Climate change information for regional impact and for risk
  assessment.

\bibitem[Raoult et~al., 2024]{aimodels}
Raoult, B., Mertes, G., and Pinault, F. (2024).
\newblock ai-models, https://github.com/ecmwf-lab/ai-models.
\newblock Accessed: 2024-06-23.

\bibitem[Rocklin, 2015]{dask}
Rocklin, M. (2015).
\newblock Dask: Parallel computation with blocked algorithms and task
  scheduling.
\newblock In Huff, K. and Bergstra, J., editors, {\em Proceedings of the 14th
  Python in Science Conference}, pages 130 -- 136.

\bibitem[Ruti et~al., 2016]{ruti2016med}
Ruti, P.~M., Somot, S., Giorgi, F., Dubois, C., Flaounas, E., Obermann, A.,
  Dell’Aquila, A., Pisacane, G., Harzallah, A., Lombardi, E., et~al. (2016).
\newblock {MED-CORDEX} initiative for mediterranean climate studies.
\newblock {\em Bulletin of the American Meteorological Society},
  97(7):1187--1208.

\bibitem[Schulzweida, 2023]{schulzweida_2023_10020800}
Schulzweida, U. (2023).
\newblock Cdo user guide.

\bibitem[Semmler et~al., 2020]{semmler2020simulations}
Semmler, T., Danilov, S., Gierz, P., Goessling, H.~F., Hegewald, J., Hinrichs,
  C., Koldunov, N., Khosravi, N., Mu, L., Rackow, T., et~al. (2020).
\newblock Simulations for {CMIP6} with the {AWI} climate model {AWI-CM-1-1}.
\newblock {\em Journal of Advances in Modeling Earth Systems},
  12(9):e2019MS002009.

\bibitem[Semmler et~al., 2021]{semmler2021ocean}
Semmler, T., Jungclaus, J., Danek, C., Goessling, H.~F., Koldunov, N.~V.,
  Rackow, T., and Sidorenko, D. (2021).
\newblock Ocean model formulation influences transient climate response.
\newblock {\em Journal of Geophysical Research: Oceans}, 126(12):e2021JC017633.

\bibitem[Sidorenko et~al., 2015]{sidorenko2015awicm}
Sidorenko, D., Rackow, T., Jung, T., Semmler, T., Barbi, D., Danilov, S.,
  Dethloff, K., Dorn, W., Fieg, K., G{\"o}{\ss}ling, H.~F., et~al. (2015).
\newblock Towards multi-resolution global climate modeling with
  {ECHAM6--FESOM}. {P}art {I}: model formulation and mean climate.
\newblock {\em Climate Dynamics}, 44:757--780.

\bibitem[Stengel et~al., 2020]{stengel2020adversarial}
Stengel, K., Glaws, A., Hettinger, D., and King, R.~N. (2020).
\newblock Adversarial super-resolution of climatological wind and solar data.
\newblock {\em Proceedings of the National Academy of Sciences},
  117(29):16805--16815.

\bibitem[Stevens et~al., 2013]{stevens2013atmospheric}
Stevens, B., Giorgetta, M., Esch, M., Mauritsen, T., Crueger, T., Rast, S.,
  Salzmann, M., Schmidt, H., Bader, J., Block, K., et~al. (2013).
\newblock Atmospheric component of the {MPI-M} {E}arth system model: {ECHAM6}.
\newblock {\em Journal of Advances in Modeling Earth Systems}, 5(2):146--172.

\bibitem[Tomasi et~al., 2024]{tomasi2024ai}
Tomasi, E., Franch, G., and Cristoforetti, M. (2024).
\newblock Can {AI} be enabled to dynamical downscaling? training a latent
  diffusion model to mimic km-scale {COSMO-CLM} downscaling of {ERA5} over
  {I}taly.

\bibitem[van~den Hurk et~al., 2018]{van2018match}
van~den Hurk, B., Hewitt, C., Jacob, D., Bessembinder, J., Doblas-Reyes, F.,
  and D{\"o}scher, R. (2018).
\newblock The match between climate services demands and {E}arth {S}ystem
  {M}odels supplies.
\newblock {\em Climate services}, 12:59--63.

\bibitem[Virtanen et~al., 2020]{2020SciPy-NMeth}
Virtanen, P., Gommers, R., Oliphant, T.~E., Haberland, M., Reddy, T.,
  Cournapeau, D., Burovski, E., Peterson, P., Weckesser, W., Bright, J., {van
  der Walt}, S.~J., Brett, M., Wilson, J., Millman, K.~J., Mayorov, N., Nelson,
  A. R.~J., Jones, E., Kern, R., Larson, E., Carey, C.~J., Polat, {\.I}., Feng,
  Y., Moore, E.~W., {VanderPlas}, J., Laxalde, D., Perktold, J., Cimrman, R.,
  Henriksen, I., Quintero, E.~A., Harris, C.~R., Archibald, A.~M., Ribeiro,
  A.~H., Pedregosa, F., {van Mulbregt}, P., and {SciPy 1.0 Contributors}
  (2020).
\newblock {{SciPy} 1.0: Fundamental Algorithms for Scientific Computing in
  Python}.
\newblock {\em Nature Methods}, 17:261--272.

\bibitem[Wang et~al., 2021]{wang2021fast}
Wang, J., Liu, Z., Foster, I., Chang, W., Kettimuthu, R., and Kotamarthi, V.~R.
  (2021).
\newblock Fast and accurate learned multiresolution dynamical downscaling for
  precipitation.
\newblock {\em Geoscientific Model Development}, 14(10):6355--6372.

\bibitem[Wang et~al., 2014]{wang2014finite}
Wang, Q., Danilov, S., Sidorenko, D., Timmermann, R., Wekerle, C., Wang, X.,
  Jung, T., and Schr{\"o}ter, J. (2014).
\newblock {The Finite Element Sea Ice-Ocean Model (FESOM) v. 1.4: formulation
  of an ocean general circulation model}.
\newblock {\em Geoscientific Model Development}, 7(2):663--693.

\bibitem[Wieners et~al., 2019a]{mpi_scenario}
Wieners, K.-H., Giorgetta, M., Jungclaus, J., Reick, C., Esch, M., Bittner, M.,
  Gayler, V., Haak, H., de~Vrese, P., Raddatz, T., Mauritsen, T., von Storch,
  J.-S., Behrens, J., Brovkin, V., Claussen, M., Crueger, T., Fast, I.,
  Fiedler, S., Hagemann, S., Hohenegger, C., Jahns, T., Kloster, S., Kinne, S.,
  Lasslop, G., Kornblueh, L., Marotzke, J., Matei, D., Meraner, K.,
  Mikolajewicz, U., Modali, K., M\"{u}ller, W., Nabel, J., Notz, D., Peters-von
  Gehlen, K., Pincus, R., Pohlmann, H., Pongratz, J., Rast, S., Schmidt, H.,
  Schnur, R., Schulzweida, U., Six, K., Stevens, B., Voigt, A., and Roeckner,
  E. (2019a).
\newblock {MPI-M MPI-ESM1.2-LR model output prepared for CMIP6 ScenarioMIP
  SSP585}.

\bibitem[Wieners et~al., 2019b]{mpi_hist}
Wieners, K.-H., Giorgetta, M., Jungclaus, J., Reick, C., Esch, M., Bittner, M.,
  Legutke, S., Schupfner, M., Wachsmann, F., Gayler, V., Haak, H., de~Vrese,
  P., Raddatz, T., Mauritsen, T., von Storch, J.-S., Behrens, J., Brovkin, V.,
  Claussen, M., Crueger, T., Fast, I., Fiedler, S., Hagemann, S., Hohenegger,
  C., Jahns, T., Kloster, S., Kinne, S., Lasslop, G., Kornblueh, L., Marotzke,
  J., Matei, D., Meraner, K., Mikolajewicz, U., Modali, K., M\"{u}ller, W.,
  Nabel, J., Notz, D., Peters-von Gehlen, K., Pincus, R., Pohlmann, H.,
  Pongratz, J., Rast, S., Schmidt, H., Schnur, R., Schulzweida, U., Six, K.,
  Stevens, B., Voigt, A., and Roeckner, E. (2019b).
\newblock {MPI-M} {MPI-ESM1.2-LR} model output prepared for {CMIP6 CMIP}
  historical.

\end{thebibliography}





\end{document}